\documentclass[proof]{pasj01}
\usepackage{lineno,graphicx,subcaption}

\Received{$\langle$reception date$\rangle$}
\Accepted{$\langle$acception date$\rangle$}
\Published{$\langle$publication date$\rangle$}

\begin{document}

\title{Magnetic Field Structure of the Crab Pulsar Wind Nebula Revealed with IXPE}
\author{Tsunefumi \textsc{Mizuno},\altaffilmark{1,}$^{*}$ Hiroshi \textsc{Ohno},\altaffilmark{2} 
Eri \textsc{Watanabe},\altaffilmark{3} Niccol\`{o} \textsc{Bucciantini},\altaffilmark{4,5,6} 
Shuichi \textsc{Gunji},\altaffilmark{3} Sinpei \textsc{Shibata},\altaffilmark{3} 
Patrick \textsc{Slane},\altaffilmark{7} and Martin C. \textsc{Weisskopf}\altaffilmark{8}}%
\altaffiltext{1}{Hiroshima Astrophysical Science Center, Hiroshima University;
1-3-1 Kagamiyama, Higashi-Hiroshima, Hiroshima 739-8526, Japan}
\altaffiltext{2}{Tohoku Bunkyo College; 
515 Katayachi, Yamagata 990-2316, Japan}
\altaffiltext{3}{Faculty of Science, Yamagata University; 
1-4-12 Kojirakawa-machi, Yamagata-shi, Yamagata 990-8560, Japan}
\altaffiltext{4}{INAF - Osservatorio Astrofisico di Arcetri;
Largo Enrico Fermi 5, 50125 Firenze, Italy.}
\altaffiltext{5}{Dipartimento di Fisica \& Astronomia, Universit\`{a} di Firenze;
Via Sansone 1, 50019 Sesto Fiorentino (FI), Italy}
\altaffiltext{6}{INAF - Sezione di Firenze;
Via Sansone 1, 50019 Sesto Fiorentino (FI), Italy.}
\altaffiltext{7}{Center for Astrophysics \textbar \ Harvard \& Smithonian;
60 Garden St, Cambridge, MA 02138, USA}
\altaffiltext{8}{NASA Marshall Space Flight Center;
Huntsville, AL 35812, USA.}
\email{mizuno@astro.hiroshima-u.ac.jp}


\maketitle

\begin{abstract}
We report a detailed study of the magnetic-field structure of the Crab pulsar wind nebula, 
using the X-ray polarization data in 2--8~keV  obtained with the Imaging X-ray Polarimetry Explorer.
Contamination of the pulsar emission to the data of the nebula region was removed  through  application of a stringent pulsation phase-cut, extracting
a phase range of 0.7--1.0 only.
We found that the electric field vector polarization angle (PA) was about $130^{\circ}$ from north to east
with the polarization degree (PD) of about 25\% at the pulsar position,
indicating that the direction of the toroidal magnetic field  
is perpendicular to the pulsar spin axis in the region close to the termination shock.
The PA gradually 
 deviated from the angle as an increasing function of the distance from the pulsar.  
There was a  region of a low PD to the west of the X-ray torus. Although such a region is expected to be located at the torus edge, where
geometrical depolarization due to a steep spatial variation of the PA is expected,
 the observed low-PD region positionally deviated from the edge.
We found that the region of low PD positionally coincided with a dense filament seen in the optical band, 
and conjecture that the low-PD region  may be produced
 through  deflection of the pulsar wind. By comparing the values of the PD at the pulsar position between the data and a model,
in which toroidal and turbulent magnetic fields  were considered,
we estimated the fractional energy of the turbulent magnetic field to be about $2/3$ of the total.
 We also evaluated a potential polarization of the northern jet    in the nebula and derived the PD and PA  to be 
about $30\%$ and $120^{\circ}$, respectively.
\end{abstract}


\section{Introduction}

The Crab Nebula complex originated from
a supernova in 1054 (SN~1054). It consists of a pulsar (PSR)
(PSR~B0531+21 or PSR~J0534+2200) and a pulsar wind nebula (PWN) powered by the PSR. 
Its short distance of about 2~kpc from the Earth and the high power of the PSR
(spin-down luminosity of ${\sim}5 \times 10^{38}~{\rm erg~s^{-1}}$) render the Crab Nebula the
apparently brightest PWN at most wavelengths.
It has been observed across all wavelengths from the radio, optical,  X-ray, to $\gamma$-ray bands in imaging, photometry, spectroscopy, and polarimetry (see\ for a review, e.g., \cite{Hester2008} and \cite{Buhler2014}).  
The Crab PWN has a broadband non-thermal spectrum from radio to $\gamma$-rays  from synchrotron
emission from accelerated electrons (and positrons) up to a few hundred MeV and from inverse Compton emission above this energy.
Therefore, it is one of the best targets to study the physics of the relativistic outflows 
(the typical Lorentz factor of the pulsar wind can be as high as ${\sim}10^{6}$) and particle acceleration.
 Whereas the  apparent size of the Crab PWN is $6^{'} \times 4^{'}$ in the optical band,
 it is roughly half of this in the X-ray band, presumably due to the strong synchrotron cooling of high-energy electrons.
Chandra observations \citep{Weisskopf2000} revealed a set of well-developed axisymmetric structures 
known as jets, inner ring, and (outer) torus, and collectively referred as ``jet-torus''.
The ring and torus are presumably produced by the combination of the nebula-wide toroidal magnetic field 
injected by the pulsar wind, and the flow pattern downstream of the wind's termination shock 
(that is considered to be traced by the inner ring).

Polarization measurements are a key to understand the magnetic-field structure and thus the interaction of the pulsar wind with the
ambient medium.
In particular, particles responsible for
X-ray and $\gamma$-ray emission suffer from severe synchrotron cooling
(in a given magnetic field the cooling time is inversely proportional to the electron energy),
hence X-ray and $\gamma$-ray polarimetry 
is a valuable probe of the magnetic-field structure close to where particles are accelerated.
High-resolution optical polarimetry observations of the Crab PWN 
revealed high polarizations 
associated with a few features known an ``wisps'' and a ``knot'' close to the PSR 
on top of a nebula-wide polarization pattern 
that broadly runs in the east-west directions (\cite{Moran2013}, \cite{Hester2008}). 
In X-rays, OSO-8 detected significant off-pulse (OP) polarizations with a polarization degree (PD) of ${\sim}$20\% and 
electric field vector polarization angle
(PA) of ${\sim}155^{\circ}$ (measured on the plane of the sky from north to east) in soft X-rays (${\le}$10~keV), establishing that synchrotron processes dominate the emission \citep{OSO8}.
Subsequent studies in soft/hard X-ray and $\gamma$-ray bands reported similar PDs in the OP phase 
(\cite{PolarLight}, \cite{PoGO+}, \cite{AstroSat}, \cite{SPI}, \cite{IBIS}).
Although these measurements could in principle provide vital information about the magnetic field properties close to where
the particle acceleration occurs, their detections were marginal (${\le}4\sigma$ level) except for the one with OSO-8,
and the results lacked any spatial information.

In 2022, the Crab PSR and PWN were observed by the Imaging X-ray Polarimetry Explorer, IXPE \citep{IXPE-Crab}, 
the first mission devoted to spatially-resolved polarization measurements in X-rays \citep{IXPE-mission}. 
The IXPE observation provided us with much better PSR and PWN polarization measurements of the Crab PSR/PWN than any of the past X-ray observatories.
\citet{IXPE-Crab} reported with IXPE the first X-ray detection of a significant polarization with a PD of ${\sim}$15\%  from the PSR emission
only in the core of the main pulse, while the total pulsed emission of the PSR was 
consistent with being unpolarized.
The strong upper limit of PD ${\sim}$6\% at the 99\% confidence level is inferred from the reported Stokes parameters.

 They also revealed the first X-ray polarization map of the PWN, which
 shows the toroidal magnetic field structure close to the PSR position.
The
PD distribution  was  highly asymmetric about the projected torus axis.

Here we report the results of in-depth analyses of the IXPE observation data of the Crab PWN.
Whereas the initial results of the Crab PWN (and PSR) observation were reoprted by
\citet{IXPE-Crab}, detailed studies of the nebula's magnetic field properties were deferred to following works.
For example, how the magnetic field direction and turbulence develop in the nebula and the relation with known structures are yet to be
examined in detail.
Here we focus on the PWN and investigate the polarization (magnetic field) properties in detail.
This paper is structured as follows. Section~2 describes the observations and data reduction.
Then we describe  our polarization analysis in section~3
and discuss the obtained polarization properties in section~4.
Finally, the summary is given in section~5.

\section{Observations and Data Reduction}
IXPE was  launched on 2021 December 9. Since then, it has provided us with 
new insight into almost all classes of X-ray objects (PWNe/PSRs, black-hole binaries, active galactic nuclei, etc.), thanks to its 
advanced capabilities in imaging, photometry, spectroscopy, and polarimetry.  IXPE consists of three polarization-sensitive
gas-pixel detectors (\cite{Costa2001}, \cite{Baldini2021}), 
placed at the focal planes of three sets of Wolter-1 mirror module assembly \citep{Soffitia2021}.
An X-ray photon focused by the mirror and absorbed by the gas in the detector
ejects a photoelectron, most likely in the direction of  the electric vector, and subsequently produces a charge cloud.
The PD and PA of a polarized source are determined from the
angular distribution of the tracks made by the initial photoelectron.
Each pair of the mirror and detector is named detector unit (DU) 1, 2,  or 3.
The mirrors give angular resolutions of ${\le}30^{''}$ in\ half-power diameter (HPD) and a field of view of  
a $\timeform{12.9'} \times \timeform{12.9'}$
square, which are adequate for spatially resolving the Crab PWN.

The mirrors have an on-axis effective area of 590~${\rm cm^{2}}$ at 4.5~keV 
(calibrated using Ti-K)
for the three telescopes combined.
The focal-plane detectors are sensitive to polarization in X-rays from 2 to 8~keV,
and reduce the effective area to 26~${\rm cm^{2}}$ at 4.5~keV and give
the peak effective area of 80~${\rm cm^{2}}$ at 2.3~keV (calibrated using Mo-L)
if the detector quantum efficiency
and event reconstruction efficiencies are taken into account.
Although the net effective area  is a decreasing function of energy
and falls to 2.3~${\rm cm^{2}}$ at 8~keV, the polarization sensitivity improves with  energy.
The modulation factor $\mu$, defined as the degree of modulation of the initial directions of the photoelectrons
for a 100\%-polarized source,
is ${\sim}15$\% at 2~keV and  ${\ge}$50\% at 8~keV.

The Crab PSR/PWN was observed twice with IXPE in 2022 between February 21 and March 7 for a total on-source time of
${\sim}92~{\rm ks}$. We also performed a simultaneous Chandra observation of the Crab (ObsID 23539).
We used the same datasets as those in \citet{IXPE-Crab}, where
the following corrections to the publicly available level-2 event files of IXPE were applied:
(1)  the energy correction to compensate for the time-dependent charge-to-energy conversion,  using onboard calibration sources
\citep{Ferrazzoli2020},
(2)  a World Coordinate System (WCS) correction to account for the slight offset among the three DUs,
(3)  an aspect-solution correction to remove  spurious offsets in the pointing solution
(due to  transitions between different star trackers activated in turn  in orbit), and
(4)  the barycenter correction, using the most recent optical coordinates and the ICRS reference frame. 
For details of the corrections,
see \citet{IXPE-Crab}.

To study the PWN, we also applied a pulse phase cut. Following \citet{IXPE-Crab}, we defined the OP phase as 
0.7--1.0 (with the main X-ray pulsar peak at phase 0.13).
We also examined events in the OP plus bridge phase 
(phase range of 0.2--0.4), but found that
the count rate and polarization properties of the PWN in the vicinity (${\sim}15^{"}$) of the PSR were significantly affected 
by the contaminating signals from the PSR.
We, therefore, analyzed only the data of the OP phase in the following analysis.
\section{Data Analysis}
\subsection{Polarization Map of the Crab PWN}

We  analyzed the level-2 event list of IXPE with the corrections applied and the 
contamination of the PSR emission removed (see section~2) in the FITS format \citep{IXPE-SOC}.
The reduced FITS file in a table format contains columns  of values of the Stokes parameters
$q_{k} \equiv 2 \cos2\phi_{k}$ and $u_{k} \equiv 2 \sin2\phi_{k}$,  where $\phi_{k}$  is the reconstructed photoelectron direction 
of the event number $k$, in addition to columns of time, energy channel, and sky and detector coordinates
commonly used in X-ray data.
The event-by-event Stokes parameters can now be used to determine the polarization integrated over energy and/or position and/or pulse phase
(see  \citet{Kislat2015} and \citet{Vink2018}).      

Denoting the modulation factor and effective area as $\mu_{k}$ and $A_{k}$, respectively, the
weighted Stokes parameters $I$, $Q$, and $U$  are calculated as
$I=\Sigma 1/A_{k}$, $Q=\Sigma q_{k}/\mu_{k}/A_{k}$, and $U=\Sigma u_{k}/\mu_{k}/A_{k}$, respectively, and the variances, as
$V(Q)=\Sigma (1/\mu_{k})^{2}(1/A_{k})^2 \cdot \Sigma q_{k}^{2}/\Sigma 1$
and $V(U)=\Sigma (1/\mu_{k})^{2}(1/A_{k})^2 \cdot \Sigma u_{k}^{2}/\Sigma 1$.
Source polarization is calculated as
${\rm PD}=\sqrt{Q^{2}+U^{2}}/I$ and ${\rm PA}=\frac{1}{2}\arctan(U/Q)$,
and their errors are estimated through the propagation of errors in
$Q$, $U$, and $I$. 
See the Appendix~1 for detail.

In our formulations, all relevant parameters ($I$, $Q$, $U$, $V(Q)$, and $V(U)$) are additive quantities.
Therefore, if we prepare these maps in fine resolution, we can always apply arbitrary binning and 
evaluate PD, PA, and the errors of each binned pixel.
Our formulations are slightly different from those in
the standard software \texttt{ixpeobssim} \citep{Baldini2022}, which
is a simulation and analysis framework specifically developed for the IXPE data.
While \texttt{ixpeobssim} calculates errors of Stokes parameters, PD, and PA adequately if the number of events is sufficiently large
in each pixel, it does not allow flexible binning of fine-resolution maps. We, therefore, adopt our formulations in this study.
Before beginning the detailed polarization analysis for the present work, 
we validated our formulations by comparing the result with the source
polarization value (integrated over the ellipse that delineates the spine of the X-ray torus of \citet{Ng2004}; see figure~1) 
obtained  with the \texttt{ixpeobssim} of version 28.4.0.
As a result, we confirmed that the  PD and PA derived with the two methods were almost identical,
with the differences much smaller than the (already small) statistical errors;
our formulations and \texttt{ixpeobssim} give PD of $(23.2 \pm 0.6)\%$ and $(23.3 \pm 0.6)\%$, respectively,
and PA of $(137.7\pm0.7)^{\circ}$ and $(137.6\pm0.7)^{\circ}$, respectively.

We  began our analysis with maps of $I$, $Q$, $U$, etc. in fine binning with a\ pixel size  of 
$\timeform{2.6"} \times \timeform{2.6"}$.
Before constructing  the maps, we need to apply one more correction known as the leakage correction \citep{PolLeak}.
    We adopted the linear  formula for the expected leakage distribution  described in \citet{PolLeak},
and calculated the leakage maps of normalized
$Q$ and $U$ (denoted as $Q_{\rm N, leak}$ and $U_{\rm N, leak}$, respectively), using the Chandra image of the Crab (figure~1(a)).
Then we derived the leakage-corrected Stokes parameters  according to
$Q_{\rm lcor}=Q-I \cdot Q_{\rm N, leak}$ and $U_{\rm lcor}=U-I \cdot U_{\rm N, leak}$.
Figures~1(b) and (c) show the obtained leakage-corrected $I$ and PD maps of the Crab PWN in 2--8~keV.
There, we applied sliding-box smoothing with $5 \times 5$ pixels to increase the photon statistics while maintaining the original resolution,
in order to investigate the polarization properties in detail.
Although one may bin maps into larger pixels, the procedure
may introduce depolarization due to a mismatch between the pixel locations and polarization distributions.
We instead applied sliding-box smoothing 
to avoid such a potential drawback.
For completeness, we also show maps with normal binning in Appendix~2.
The toroidal magnetic field structure was clearly visible around the PSR
in the figure, as reported by \citet{IXPE-Crab}.  
In addition, we  found  a few notable properties  about the Crab PWN polarization as follows. 
\begin{enumerate}
\item The magnetic-field direction slightly deviates from that of the torus major axis (see also \cite{IXPE-Crab}), 
the tendency of which is particularly evident outside of the ellipse that delineates the torus but already present inside.
\item There are high PD areas (PD${\ge}40\%$) 
in the north and south of the torus
(already pointed out by \cite{IXPE-Crab}).
\item Regions of a very-low PD (blue/black regions in the figure 1(c)) are  identified  to the east and west of the torus.
The low-PD area  to the west of the torus does not coincide with the southwest edge of the torus.
\item In the north and south of the torus, moderately low PD  regions (red regions in the figure~1(c)) 
positionally coincide with 
the north/south jets seen in the Chandra image.
\end{enumerate}

\begin{figure}
 \begin{tabular}{cc}
  \begin{minipage}{0.5\textwidth}
   \centering
   \includegraphics[width=\textwidth]{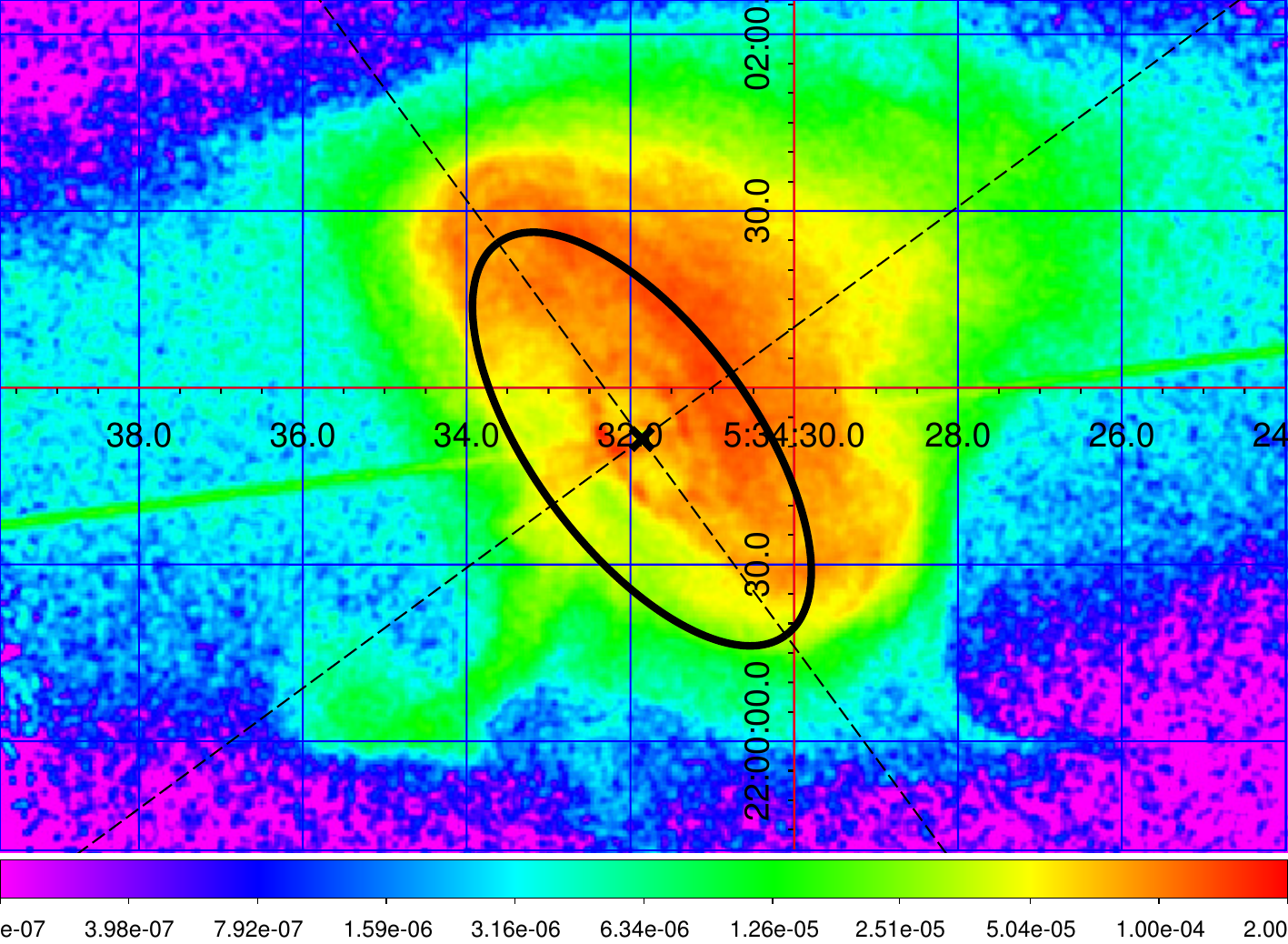}
   \subcaption{}
  \end{minipage}
  \begin{minipage}{0.5\textwidth}
   \centering
   \includegraphics[width=\textwidth]{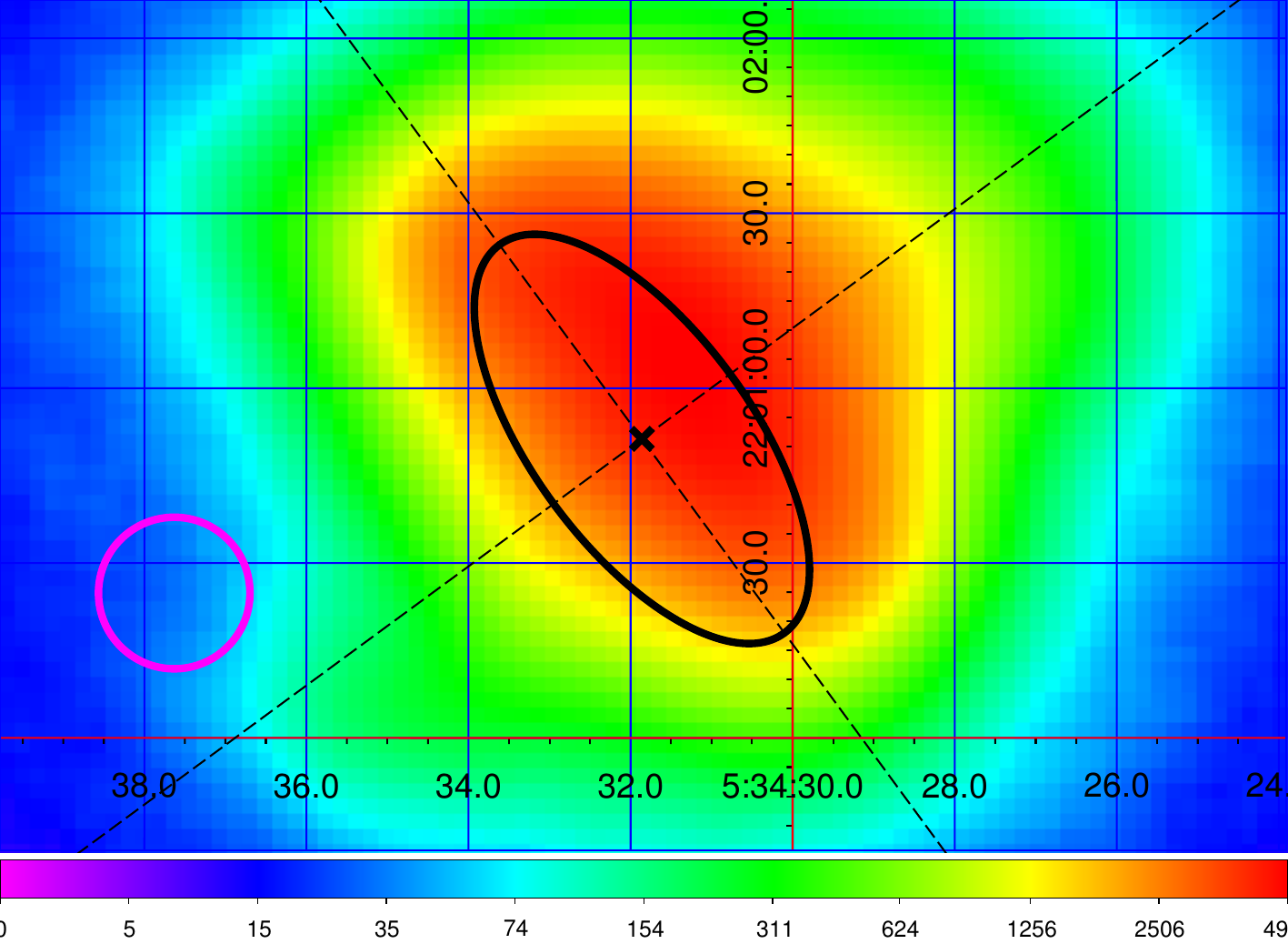}
   \subcaption{}
  \end{minipage} \\
  \begin{minipage}{0.5\textwidth}
   \centering
   \includegraphics[width=\textwidth]{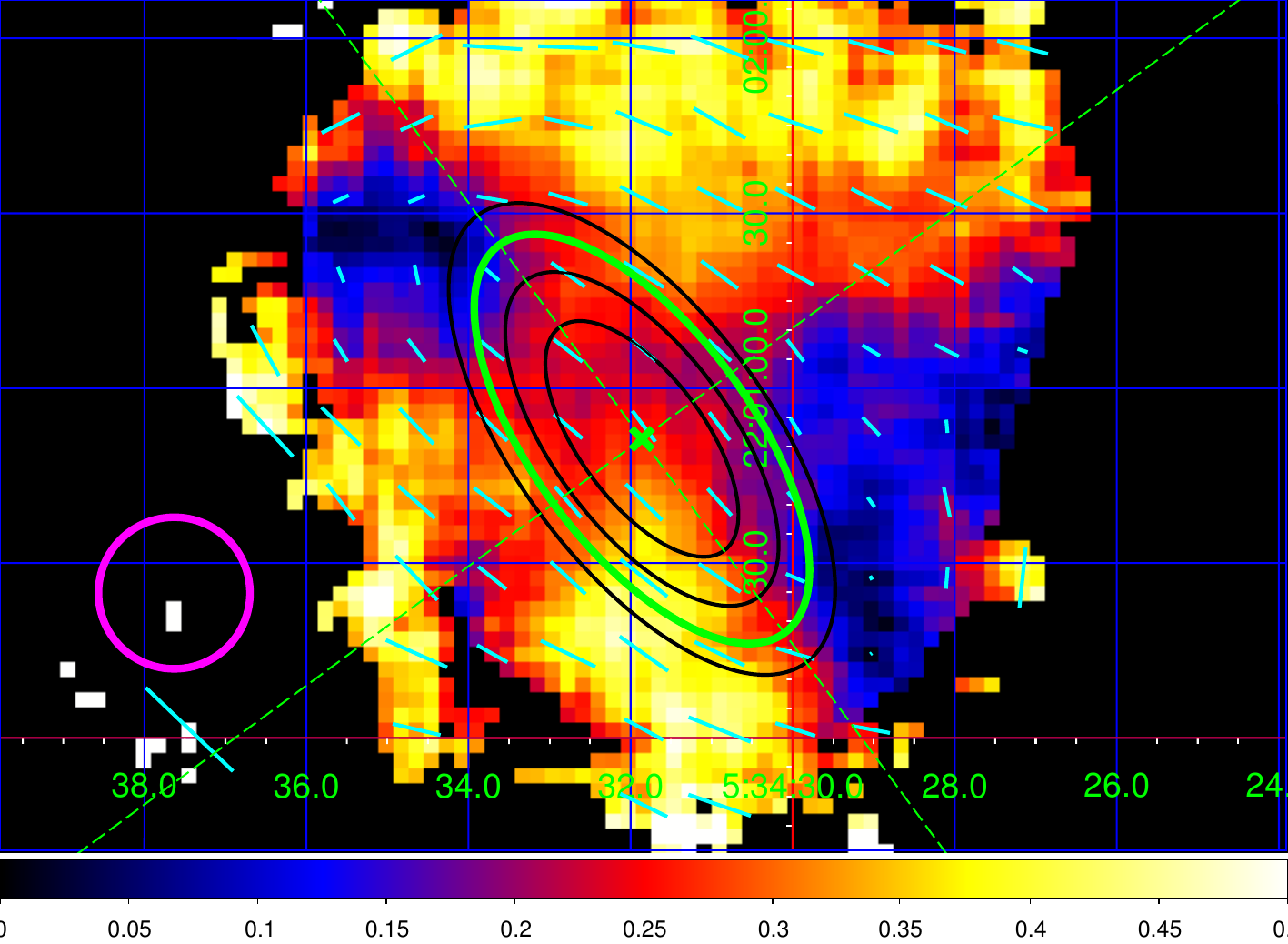}
   \subcaption{}
  \end{minipage}
 \end{tabular}
 \caption{
\small{(a)  Chandra image (ObsID 23539) of the Crab PWN (in unit of ${\rm counts~s^{-1}~cm^{-2}}$), and the
(b)  smoothed Stokes $I$ map (in unit of counts) and (c)  PD map of the Crab PWN  obtained with IXPE,
all in 2--8~keV.
Coordinates are given in right ascension (horizontal axis) and declination (vertical axis) at J2000.0.
For the PD map, only pixels with a polarization detection significance  of higher than 3$\sigma$ and ${\rm MDP_{99}<100\%}$ or those
with ${\rm MDP_{99}}<20\%$ are shown, where ${\rm MDP_{99}}$ is the minimum detectable polarization
at a 99\% confidence level.
The PSR position (cross), the spine of the X-ray torus (thick-line ellipse; in black for panels a and b but in green for panel c),
and its axes (dashed lines whose position angles are $126^{\circ}$ and $36^{\circ}$)  
according to \citet{Ng2004}, together with a magenta
circle of $26^{"}$ diameter 
(approximately the HPD of the IXPE\ image) 
are  indicated in each panel for reference. 
The PD map (panel~c) also shows
the reconstructed magnetic-field vectors (perpendicular to the PA) by segments
with a spacing of 5 pixels ($=13\arcsec$), where the
lengths are proportional to the PD.  Thin-line annular ellipses  indicate the regions used in the data analysis to investigate the positional dependence
of polarization.
We note that \citet{Ng2004} do not give the torus center position.
 Since the PSR position is very close ($\le 1\arcsec$) to the center of the inner ring \citep{Weisskopf2012}, 
 we assume that the torus center is located at the PSR position for simplicity. }
}\label{.....}
\end{figure}

\subsection{Positional Dependence of Polarization}
Here we  investigate the positional dependence of the polarization
(the first point in the notable feature list at the end of section~3.1) in detail. 
First, we   made 
the PA profiles along the major and minor axes of the X-ray torus (figure~2)  and found that the PA  was ${\sim}130^{\circ}$
at the PSR position (i.e.,\ torus center), which is close to the direction of the projected torus axis ($126^{\circ}$; \citet{Ng2004}), whereas
the PA values gradually deviated as  an increasing function of the distance from the PSR position along the major axis.
 Then, we defined a small ellipse (region~1) and three concentric annular ellipses  (regions~2--4, numbered for increasing radii)  with equal areas, 
all centered at the PSR position;
the ellipse delineating the spine of the the X-ray torus by \citet{Ng2004} is composed of regions 1, 2, and 3 (figure~1 (c)).
We calculated the PD and PA in each region by integrating the pixel values inside each region without smoothing
and summarize the results in figure~3.
We found that the PA  at the innermost region is the closest to the  angle of the projected torus axis
(that is presumably parallel to the PSR spin axis), 
and the PD and\ PA  are  decreasing and increasing functions of the distance
 from the PSR. We thus conclude that the direction of the toroidal magnetic field in the Crab PWN  
is perpendicular to the PSR spin axis in the regions close to the termination shock and
 gradually deviated to the east-west direction due to some environmental effects.  
While the physical origin of the magnetic-field deviation is not clear, 
it could be possible mismatch between the explosion symmetry axis and pulsar spin axis as discussed by
\citet{Hester2008}.

While the X-ray polarimetry probes magnetic-field direction close to the particle acceleration sites, 
the position resolution of IXPE is limited. In this regard, optical polarimetry is complementary.
Indeed, high-resolution optical polarimetry by \citet{Moran2013} gave similar results.
While structures close to the pulsar, such as the knot and wisps, have PA of ${\sim}125^{\circ}$,
the rest of the inner nebula shows the PA distribution peaked at ${\sim}165^{\circ}$.
Therefore two results provide a consistent picture of the development of magnetic field in the Crab PWN.

\begin{figure}
 \begin{tabular}{cc}
  \begin{minipage}{0.5\textwidth}
   \centering
   \includegraphics[width=\textwidth]{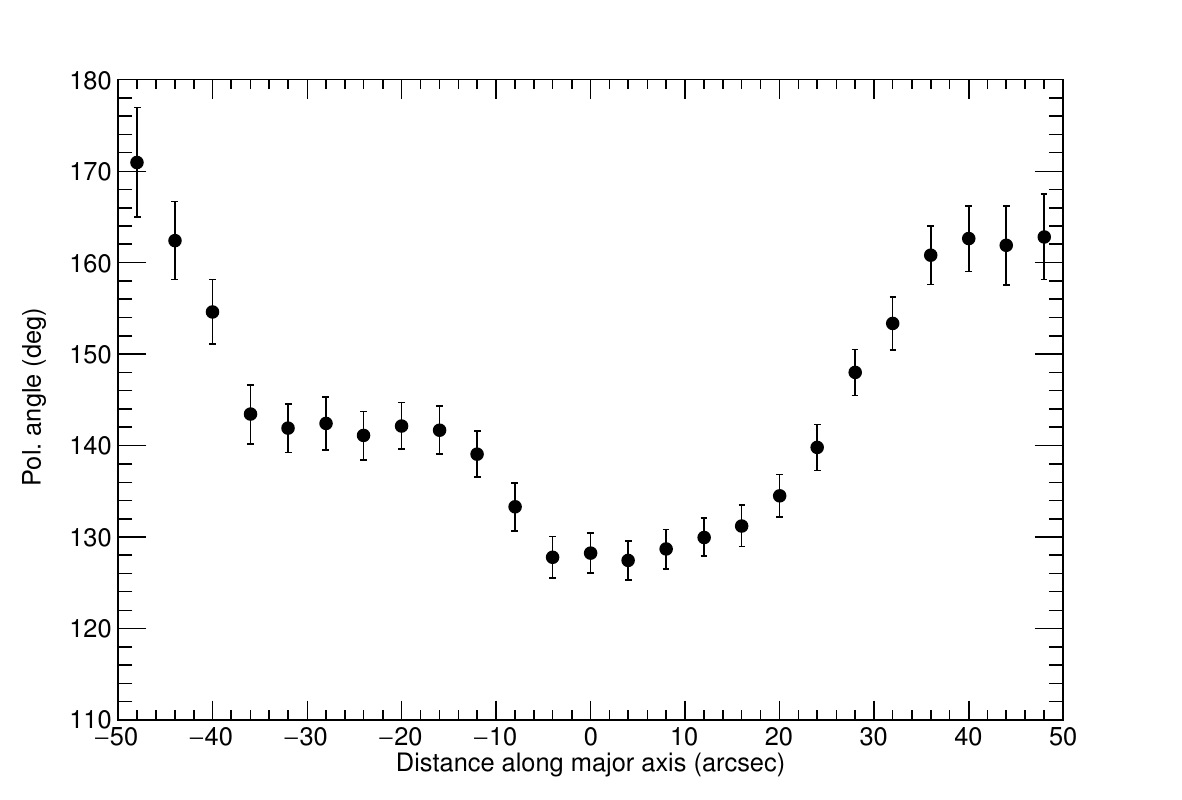}
   \subcaption{}
  \end{minipage}
  \begin{minipage}{0.5\textwidth}
   \centering
   \includegraphics[width=\textwidth]{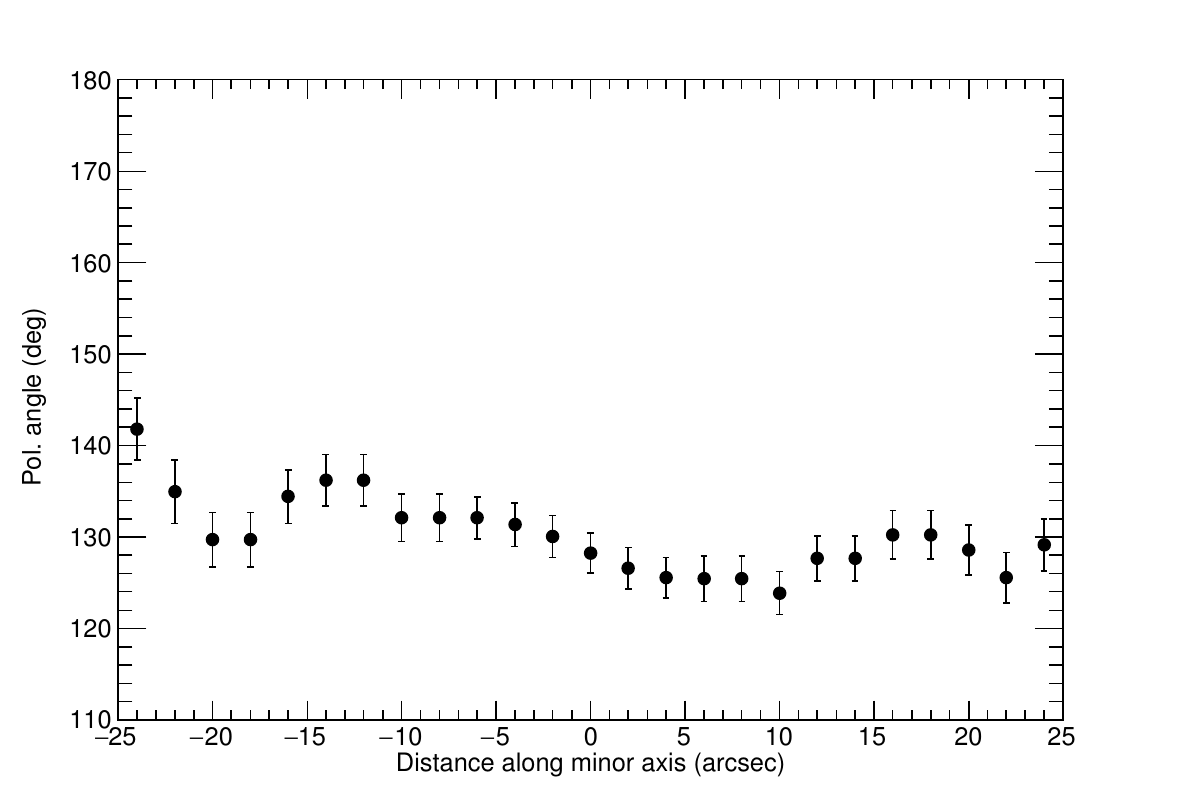}
   \subcaption{}
  \end{minipage}
 \end{tabular}
 \caption{ PA as a function of the distance from the PSR along the (a) major axis and (b) minor axis,
with  sliding-box binning ($5 \times 5$ pixels) applied.
The positive distances  are  defined as the northwest and southwest directions, respectively.
}\label{.....}
\end{figure}

\begin{figure}
 \begin{tabular}{cc}
  \begin{minipage}{0.5\textwidth}
   \centering
   \includegraphics[width=\textwidth]{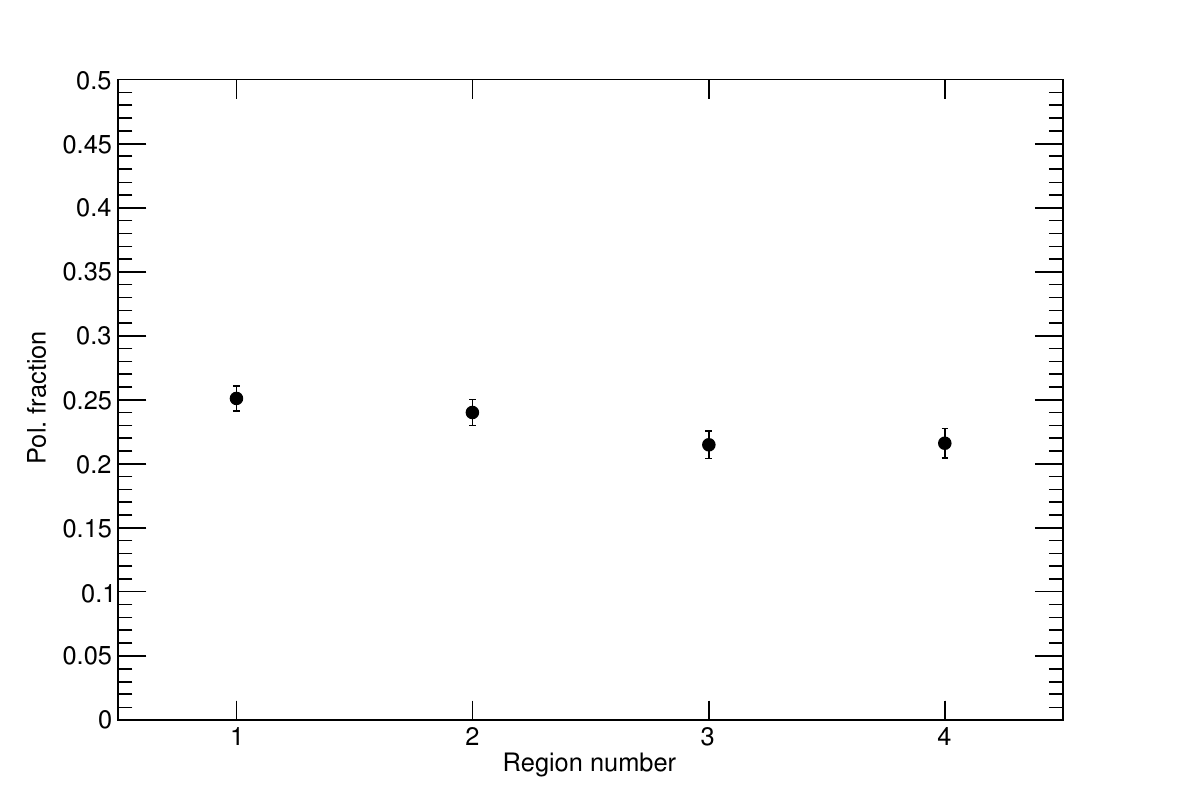}
   \subcaption{}
  \end{minipage}
  \begin{minipage}{0.5\textwidth}
   \centering
   \includegraphics[width=\textwidth]{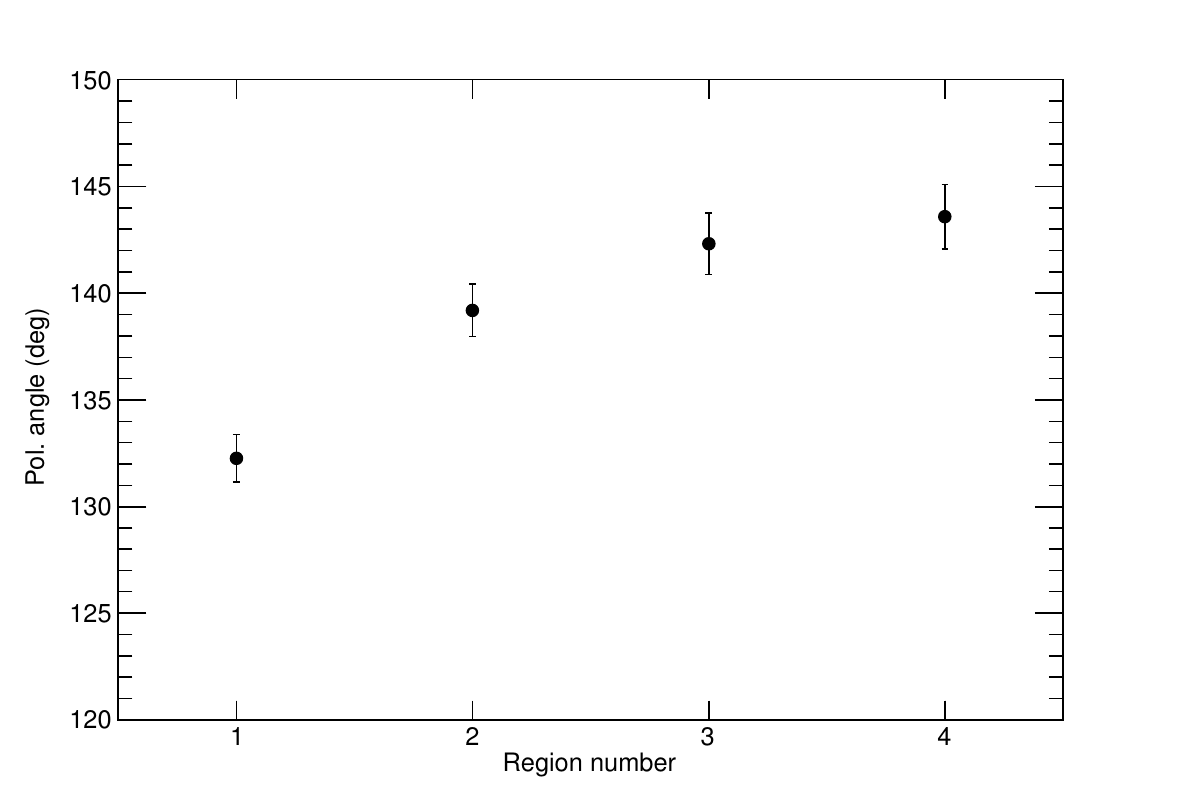}
   \subcaption{}
  \end{minipage}
 \end{tabular}
 \caption{(a) PD and (b) PA of  regions 1--4, where all the  regions have equal areas; region~1 is the closest to the PSR,
and regions~2--4  are increasingly farther. 
}\label{.....}
\end{figure}

\section{Discussion}
\subsection{Comparison with Past High-Energy Polarimetry}

X-ray and $\gamma$-ray polarimetry is a valuable probe into the magnetic-field structures of the Crab PWN (see Introduction).
Figure 4 shows the results of this work for the inner region (region 1) and for the whole nebula
integrated over the circle with a \timeform{2.5'}-radius
from \citet{IXPE-Crab}.
Also shown are reported results from other missions.
While the IXPE energy range overlaps with that of OSO-8, the obtained PA for the whole nebula was different by ${\sim}10^{\circ}$.
\citet{IXPE-Crab} argued that the apparent discrepancy of the PA could be due to the variability of the PWN, 
where structures are known to change in shape and location over a typical timescale of a few years (e.g., \cite{Schweizer2013}).
Our analysis of the same data as those used by \citet{IXPE-Crab} confirms the difference in PA,
and the argument by \citet{IXPE-Crab} about the cause of the discrepancy remains plausible.

We note that figure~4 plots the PD and PA of the whole nebula except for region~1.
Therefore, ignoring region~1, one can see that
there is a strong trend that
the PD is ${\sim}$20\% in soft X-rays and gradually increases toward the high energy.
In addition, the PA in soft X-rays is ${\sim}20^{\circ}$ offset toward the south from
the direction of the projected torus axis, but gradually approaches the direction
in the high-energy band. 
This seems reasonable as the size of the nebula is decreasing as the energy increases.
\citet{SPI} and \citet{IBIS} argued that such an energy dependence
could be due to synchrotron cooling. Higher energy electrons have a shorter lifetime.
As a result, the PA in $\gamma$-rays is expected to be perpendicular to the
magnetic-field direction near the jet and/or termination shock,
and accordingly to be parallel to the PSR spin axis and X-ray torus axis. 
The past observational data before IXPE, however, could not verify the 
hypothesis of \citet{SPI} and \citet{IBIS} due to the lack of
spatial information.
The IXPE data, for the first time, revealed how the magnetic-field structure developed from the PSR position toward the outer part of the PWN;
 the PA is close to the direction of the projected torus axis at the PSR position and gradually deviates from the angle
 as a function of the distance from the PSR (figure~2). 
The PA of region~1 is consistent with those
reported in the $\gamma$-ray band and close to the projected torus axis. Consequently, the  IXPE data  support the past speculation about how the
magnetic field changes within the nebula,
providing us with much richer information on the magnetic-field structures and thus much more solid observational evidence (see the following subsections).

\begin{figure}
 \begin{tabular}{cc}
  \begin{minipage}{0.5\textwidth}
   \centering
   \includegraphics[width=\textwidth]{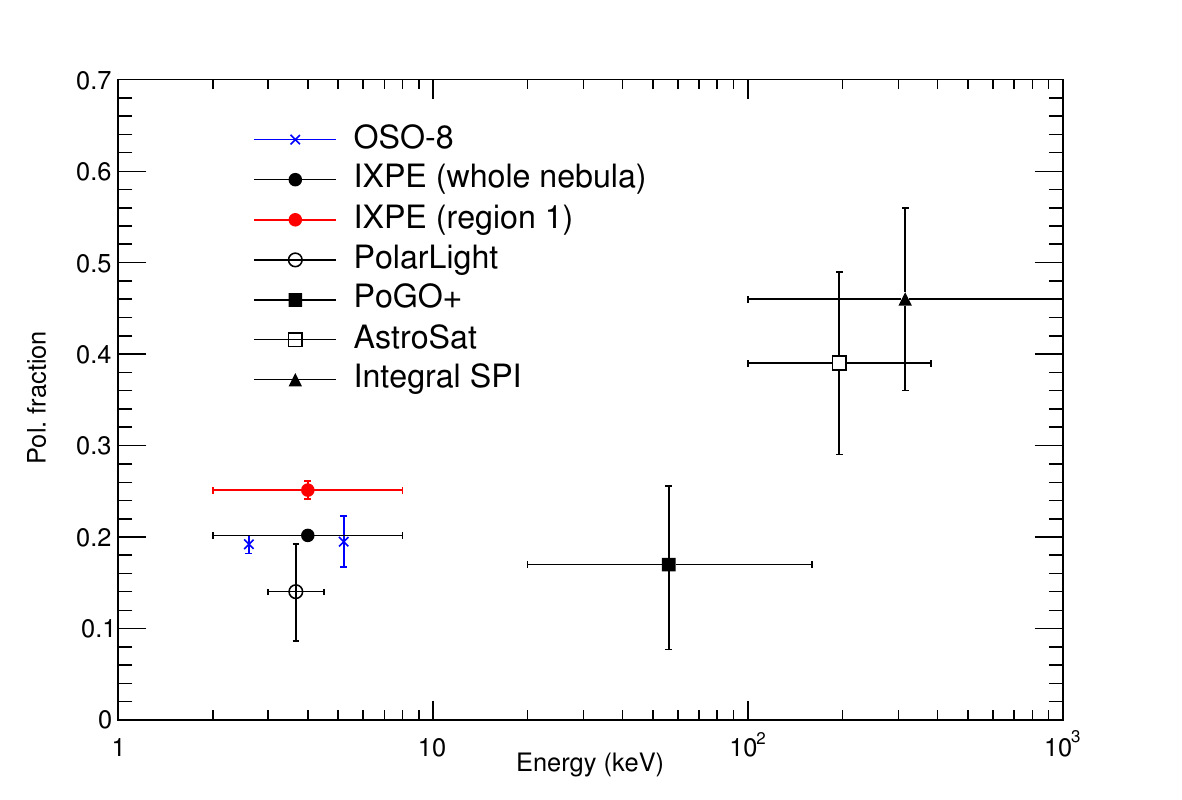}
   \subcaption{}
  \end{minipage}
  \begin{minipage}{0.5\textwidth}
   \centering
   \includegraphics[width=\textwidth]{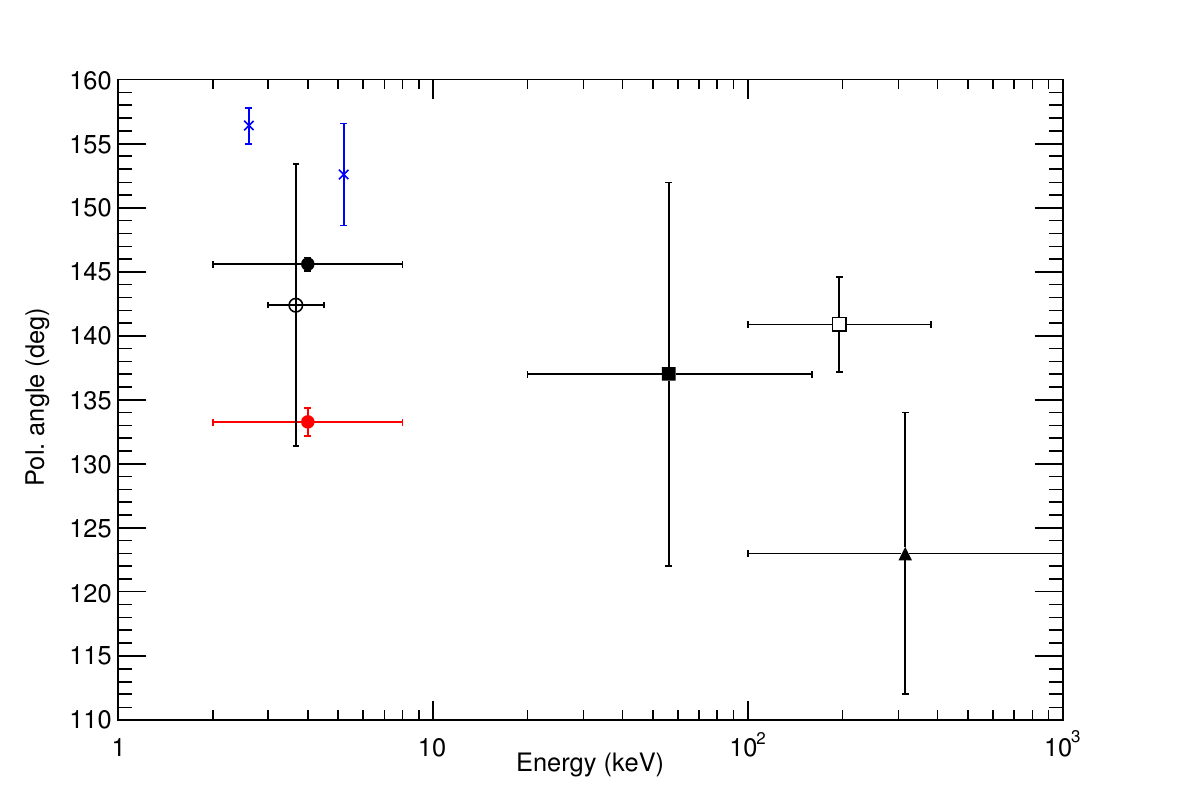}
   \subcaption{}
  \end{minipage}
 \end{tabular}
 \caption{(a) PD and (b) PA  measured with IXPE and obtained in past measurements in the X-ray and $\gamma$-ray bands. 
Note that the Integral IBIS result, which is not  plotted for clarity, gives an even larger PD and a consistent PA
compared with those by Integral SPI.
}\label{.....}
\end{figure}

\subsection{Comparison with Known Structures}

Figure~1 shows that the PD distribution is far from uniform
with the level of asymmetry much stronger than the intensity.
A steep spatial variation of the polarization within the HPD
will produce low PD areas (e.g., \cite{Nakamura2007}).
Such a geometrical depolarization, however, should still produce
a symmetric PD distribution about the projected torus axis, somewhat contrary to the observation.
As discussed in \citet{IXPE-Crab}, patchy development of the magnetic-field turbulence is a possible scenario
for a non-uniform PD distribution.
Alternatively the PD may be originally high  at high latitudes and that the emission is 
accidentally depolarized due to
environmental effects (e.g., local interaction with the ambient supernova ejecta).
A 3D simulation of the Crab PWN by \citet{Porth2014} suggests that
the toroidal magnetic field  is dominant at high latitudes except for the regions along the PSR spin 
axis. The
very high PD  observed in the Vela PWN \citep{IXPE-Vela}  is consistent with this scenario.
A possible alternating magnetic field ("striped wind model") due to an oblique rotation of the PSR
will dissipate during the propagation toward the termination shock
(e.g., \cite{Nagata2008}). Although this process will reduce the PD along the equatorial plane of the PSR,
the PD  at the high latitude will remain high.
Therefore, 
we argue that the PD of the Crab PWN was originally high
 at the high latitude but reduces due to environmental effects. 

A blue-shifted
filament in the western part of the Crab PWN
has a high column density of ${\sim}10^{21}~{\rm cm^{2}}$ (e.g., \cite{Mori2004}, \cite{Martin2021}). 
It runs  nearly parallel to the 
projected torus axis on the plane of the sky, 
and positionally coincides with a very-low PD area in the western part of the Crab PWN\footnote{
 According to figure~7 in \citet{Martin2021}, we assume that the edges of the filament are situated at
$(\alpha, \delta)_{\rm J2000.0}=$ (83\fdg625, +22\fdg008) and (83\fdg610, +22\fdg015), where
$\alpha$ and $\delta$ are right ascension and declination, respectively.}, as shown in figure~5(b). 
The filament may  alter the magnetic-field direction and produce the region of low PD. 
In addition another
low-PD area is observed at the northeast edge of the torus
(we interpret that it is also due to geometrical depolarization caused by the 
steep spatial variation of the PA).
We also note that the low-PD area and the filament in the western part of the PWN
 are positionally close to the western ``bay''-like structure at 
around right ascension $\alpha =$ \timeform{5h34m28s} (J2000.0), where there is
a sharp drop in X-ray emission, which is a 
feature of the Crab PWN commonly observed at all wavelengths from the radio to X-rays (e.g., \cite{Dubner2017} and \cite{Seward2006}).
Although the physical origin of the filament and the ``bay''
in the west of the torus is unknown,
the toroidal magnetic field in the nebula flow seems to experience significant changes of direction by the filament
(see also figure~2(a)) and another
low-PD area is produced.
Subsequently, the plasma  is blocked from flowing further westward, which decreases the X-ray emission as observed.

\begin{figure}
 \begin{tabular}{cc}
  \begin{minipage}{0.5\textwidth}
   \centering
   \includegraphics[width=\textwidth]{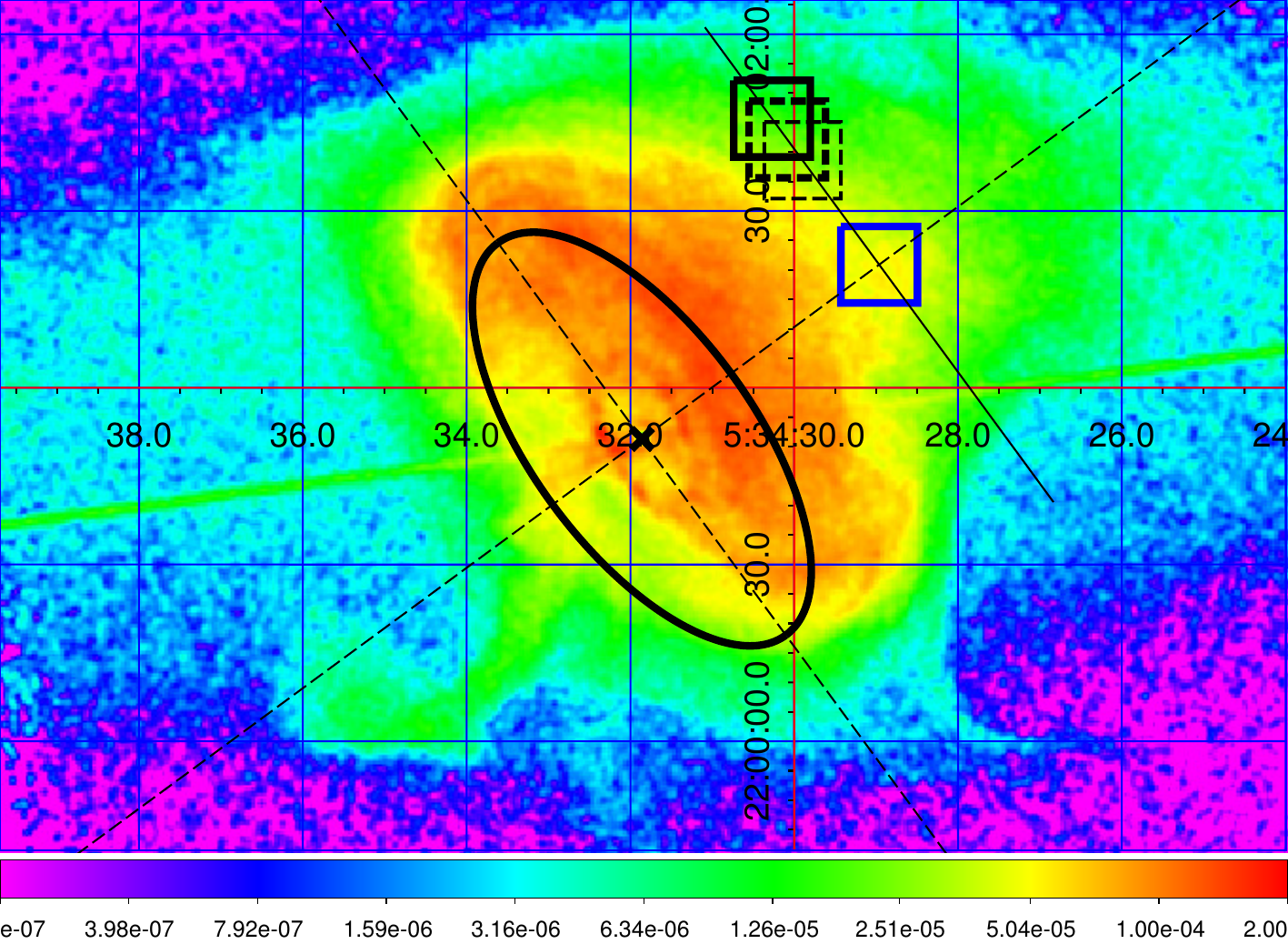}
   \subcaption{}
  \end{minipage}
  \begin{minipage}{0.5\textwidth}
   \centering
   \includegraphics[width=\textwidth]{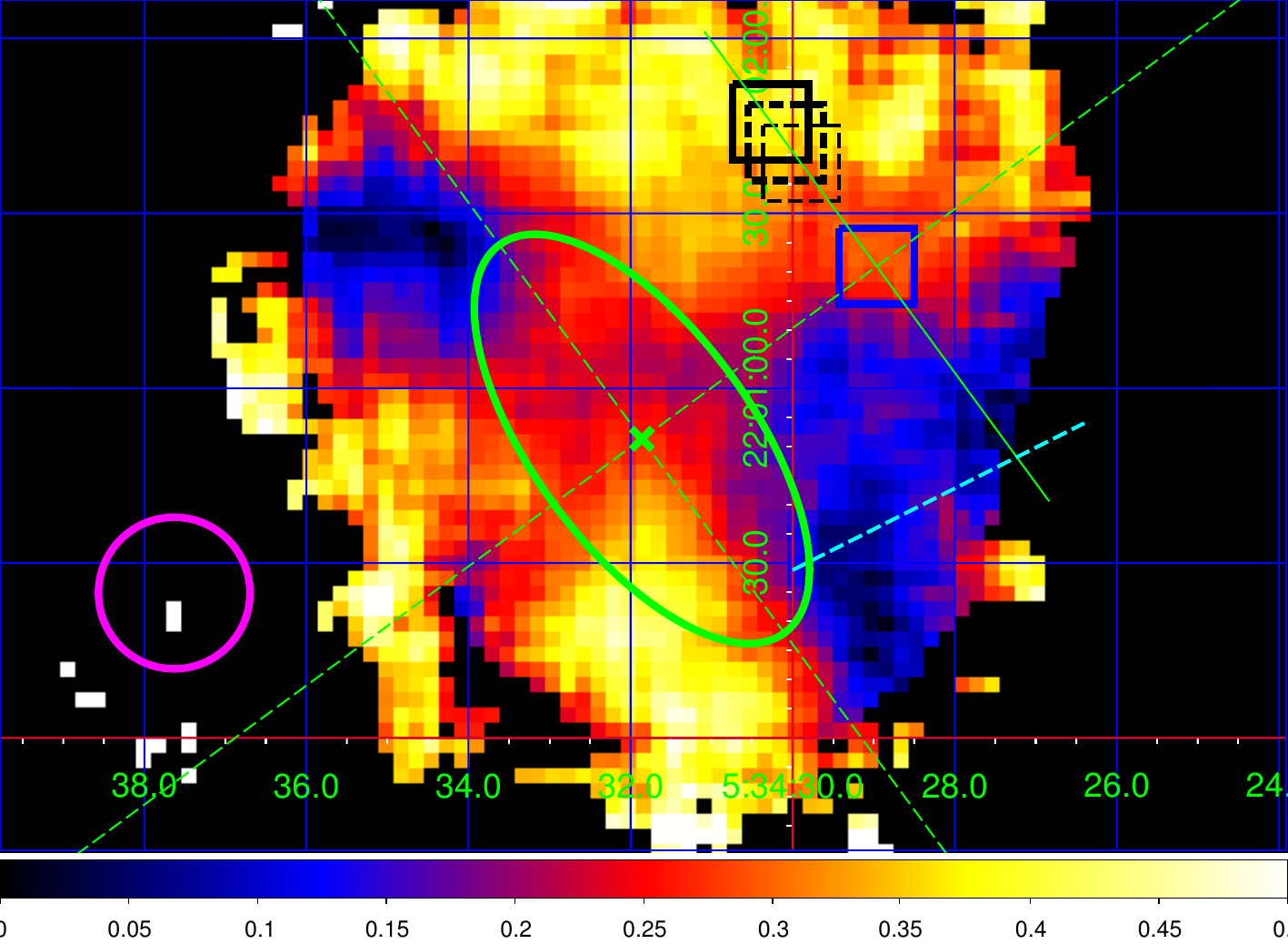}
   \subcaption{}
  \end{minipage}
 \end{tabular}
 \caption{(a)  Chandra image of the Crab PWN (in unit of ${\rm counts~s^{-1}~cm^{-2}}$) and (b) the PD map of the Crab PWN  obtained with IXPE,
both in J2000.0 equatorial coordinate.
They are basically the same as figures~1(a) and 1(c), respectively, but with the regions
used to study the polarization of the northern jet
(blue and black boxes for source and off-source regions, respectively) overlaid. See section~4.4 for details.
In panel (b), the position of a dense filament,  which is identified in the optical band and contributes to the X-ray absorption
(\cite{Mori2004}, \cite{Martin2021}), is indicated by a  cyan dashed line (section~4.2).
}\label{.....}
\end{figure}

\subsection{Comparison with a Simulation}

PWNe are bubbles of relativistic particles and  magnetic fields produced  through the interaction
 between the ultra-relativistic pulsar wind and the ambient medium. They have been extensively studied
to understand the dynamics of the pulsar wind and associated particle acceleration.
We aim to evaluate the magnetic field turbulence of the Crab PWN by comparing the observed polarization properties with a dedicated simulation.

\citet{KC84a} developed a pioneering 1D magnetohydrodynamic (MHD) model to describe the physics of the PWN, which  has been widely  accepted since. 
Here is the overview of the standard physical processes according to the model. The ultra-relativistic wind produced by the pulsar slows down at the termination shock. There,
 the toroidal magnetic field is compressed, the plasma is heated, and particles are accelerated.
As  the post-shock flow expands toward the nebula's edge, a bubble of high-energy particles and  intense magnetic field is  generated.
The key parameter to characterize the process is the $\sigma$-parameter, the ratio of the magnetic energy flux to the kinetic energy flux  
immediately before the termination shock occurs. 

Kennel \& Coroniti (\yearcite{KC84a}, \yearcite{KC84b}) suggested $\sigma \sim 0.003$ for the Crab PWN to reproduce the observed 
synchrotron luminosity and expansion velocity.
With  this value of $\sigma$, however, the flow velocity of the medium is predicted to be non-relativistic.
It hence 
cannot explain the apparent  asymmetry  in the X-ray surface brightness  along the northwest-southeast axis, which presumably  originates 
from
 Doppler boosting and relativistic aberration. In this regard, \citet{Mori2004} suggested $\sigma$ 
${\sim}$0.05 to produce the relativistic flow speed and reproduce the observed surface-brightness  asymmetry.
This much larger value of $\sigma$, however, raises a different problem, as pointed out by \citet{Shibata2003};   an intensity peak close to the termination shock, which should spatially coincide with
 the inner ring, is predicted to appear, contrary to the observed 
X-ray emission from the torus being much brighter than from the inner ring.
There is also another problem with Kennel \& Coroniti's model; i.e.,  a purely toroidal magnetic field assumed in the model 
predicts ``lip-shaped'' synchrotron emission to be generated as shown by \citet{Shibata2003},
again contrary to the  observed ``ring-like'' shape of the torus. In short, 
the model by \citet{KC84a} (hereafter  referred to as ``the KC model'')  does not fully  agree with observation
under the assumption of a purely toroidal field, regardless of the ${\sigma}$ value.

We note that magnetic field turbulence is not taken into account in the KC model,
but it is also important and has been studied extensively (e.g., \cite{Luo2020}). Indeed,
   \citet{Shibata2003} proposed a modified model from the KC model, introducing magnetic-field turbulence, 
which gives a solution to the problem mentioned above
with the comparatively large $\sigma$ of $\sim 0.05$ that is necessary to explain the surface-brightness asymmetry.
They argued that an alternating magnetic field due to an oblique rotation of the PSR will cause the dissipation of the
 magnetic field through magnetic reconnection.
 As a result, beyond the termination shock, the flow will be decelerated,
and the magnetic field will accumulate and be amplified.
 This means that $\sigma$  is in effect reasonably  small and hence the resultant luminosity intensity peak will explain the observation
(see their figure~3).
In addition, the magnetic-field turbulence can  produce the  annular-elliptical emission in agreement with the observation.

The degree of the magnetic-field turbulence can be constrained by comparing
the X-ray polarization data and the model. In this regard,
\citet{Nakamura2007}  developed a polarization-distribution model of the torus emission, assuming the canonical value of the $\sigma$-parameter (0.003)
and a flow velocity of 0.2$c$ (where $c$ is the speed of light) to reproduce the observed north-south asymmetry  
in the torus brightness.
They compared the predicted PD integrated  over the entire nebula with the OSO-8 result (PD${\sim}$20\%) and found that
the energy of the turbulent magnetic field  was ${\sim}$60\% of the total magnetic-field energy  for the randomness parameter, $b$,  of 0.6 in their definition. 
 More recently, \citet{Bucciantini2017}  developed a similar model  in which magnetic-field turbulence was considered and the inner ring and jet
 were taken into account in addition to the torus. Their result  of the degree of turbulence based on the OSO-8 result was very similar to that by \citet{Nakamura2007}.
Their best estimate of the magnetic-field fluctuation parameter was\ 0.7, giving a ratio of energies
of the turbulent and toroidal magnetic field of 3:2.

These pioneering works are based on the integrated PD and hence are subject to uncertainty. We therefore
carried out a new model calculation, convolved with the IXPE responses 
(effective area, energy resolution, spatial resolution, and modulation response), and directly compared 
the result with the IXPE observation to evaluate the magnetic-field turbulence.
In this work, we do not adopt 2D and 3D MHD simulations in the model because recently reported results with  2D and 3D MHD simulations   
of the PWN (e.g., \cite{DelZanna2006} and \cite{Porth2014}) did yield 
 broadly symmetric magnetic-field structures, in disagreement with the observation, although they successfully reproduced more detailed structures such as jets, wisps, and a knot.
Instead, we develop a phenomenological model with aid of the (1D) KC model in the manner described below. 
We aim to reproduce the overall trend of the observed properties
of the X-ray torus where non-axisymmetry of the PD is less prominent than outside
and evaluate the magnetic-field turbulence therein. 
\begin{enumerate}
\item The Stokes parameters in the flow frame and the observer frame are calculated  in the manner described in \citet{Nakamura2007}.
\item The PWN is modeled with a simple equatorial wedge (see figure~1 of \cite{Shibata2003}), 
parameterized with the inner radius $r_{\rm s}$, outer radius $r_{\rm n}$,
and semi-opening angle $\theta_{0}$. Specifically, we assume the following values:
\begin{itemize}
\item $r_{\rm s}=0.1~{\rm pc}$ to match  the observed inner ring,
\item $r_{\rm n}=0.6~{\rm pc}$ to reproduce the size of the X-ray torus,
\item $\theta_{0}$ is $\pm10^{\circ}$ to match  the observed X-ray structure, which is not spherically symmetric but torus-like,
\item flow velocity $v=0.2c$ to reproduce the observed north-south asymmetry of the X-ray torus, and
\item
wedge position and inclination angles of $126.3^{\circ}$ and $63.0^{\circ}$, respectively,  taken from \citet{Ng2004}.
\end {itemize}
\item The magnetic field is assumed to consist of a toroidal component and a turbulent (isotropic) one in the flow frame with $b=0.6$  according to
\citet{Nakamura2007} and \citet{Bucciantini2017}.
\item The radial profile of the magnetic field distribution is assumed to follow the KC model prediction
 under an assumption of a canonical value of $\sigma=0.003$.
As in \citet{Nakamura2007}, the pulsar wind luminosity and wind Lorentz factor are assumed to be $5 \times 10^{38}~{\rm erg~s^{-1}}$
and $3 \times 10^{6}$, respectively.
\item  The electron distribution within the PWN wedge is assumed to be uniform with a power-law index of 3,
 giving the index 2 for the synchrotron emission that approximates the observed spectrum in soft X-rays (e.g., \cite{CrabSpec}, \cite{Mori2004}).
We calculate the observed synchrotron emission, taking into account the flow velocity and the magnetic field described above.
\end{enumerate}

Figure~6 shows the obtained count and PD maps convolved with the IXPE response.  
The PD map shows depolarization at the edges along the torus major axis,
and the reconstructed magnetic-field direction is perpendicular to the projected torus axis.
Figure~7 summarizes the comparisons of the count and PD distributions between the data and model.
Here, since the PSF size of IXPE is moderate,
 the obtained PD and PA are inevitably affected by high-PD areas outside the torus 
and the low-PD area in the western part of the PWN.
In addition, as described in section~3.2, the magnetic field structure away from the termination shock
is likely to be affected by environmental effects.
Hence, we compare the PD between the data and model at the PSR position
where the contamination from the high/low PD areas is minimal. 
The obtained values of PD are $(27.0 \pm 2.0)\%$ and 33.5\%, respectively. With ($1-b$) scaling of the PD
as in \citet{Nakamura2007}, we obtained $b=0.68$ as our best estimate  
for the magnetic-field turbulence there. 
Although the observed  data and model broadly agree, a closer look identifies some differences; 
the model gives a wider distribution along the major axis and the peak position  closer to the PSR along the minor axis.
By reducing $r_{\rm n}$ by 0.1~pc, we obtain a better agreement in the count profiles along the major axis,
but the
agreement worsens along the minor axis. Increasing $r_{\rm n}$ gives an opposite trend.
Through these procedures the model PD and PA are hardly affected (the difference is much smaller than the statistical error of data). 
In consequence, our estimate of the magnetic-field turbulence (parameter $b$) is 
not affected against fine-tuning
the torus shape,
and revealed that the turbulent magnetic-field gives about 2/3 of the total magnetic-field energy
close to the termination shock.

We also note that the observed large non-axisymmetry of the PD distribution 
(distribution of magnetic-field turbulence), particularly outside the X-ray torus, does not agree with predictions by 2D and 3D MHD simulations.
The IXPE results thus require further development of such theoretical works. For example, 
it is important to run longer 3D simulations and examine how the
magnetic-field dissipation develops in the Crab nebula, as pointed out by \citet{Porth2014}.

\begin{figure}
 \begin{tabular}{cc}
  \begin{minipage}{0.5\textwidth}
   \centering
   \includegraphics[width=\textwidth]{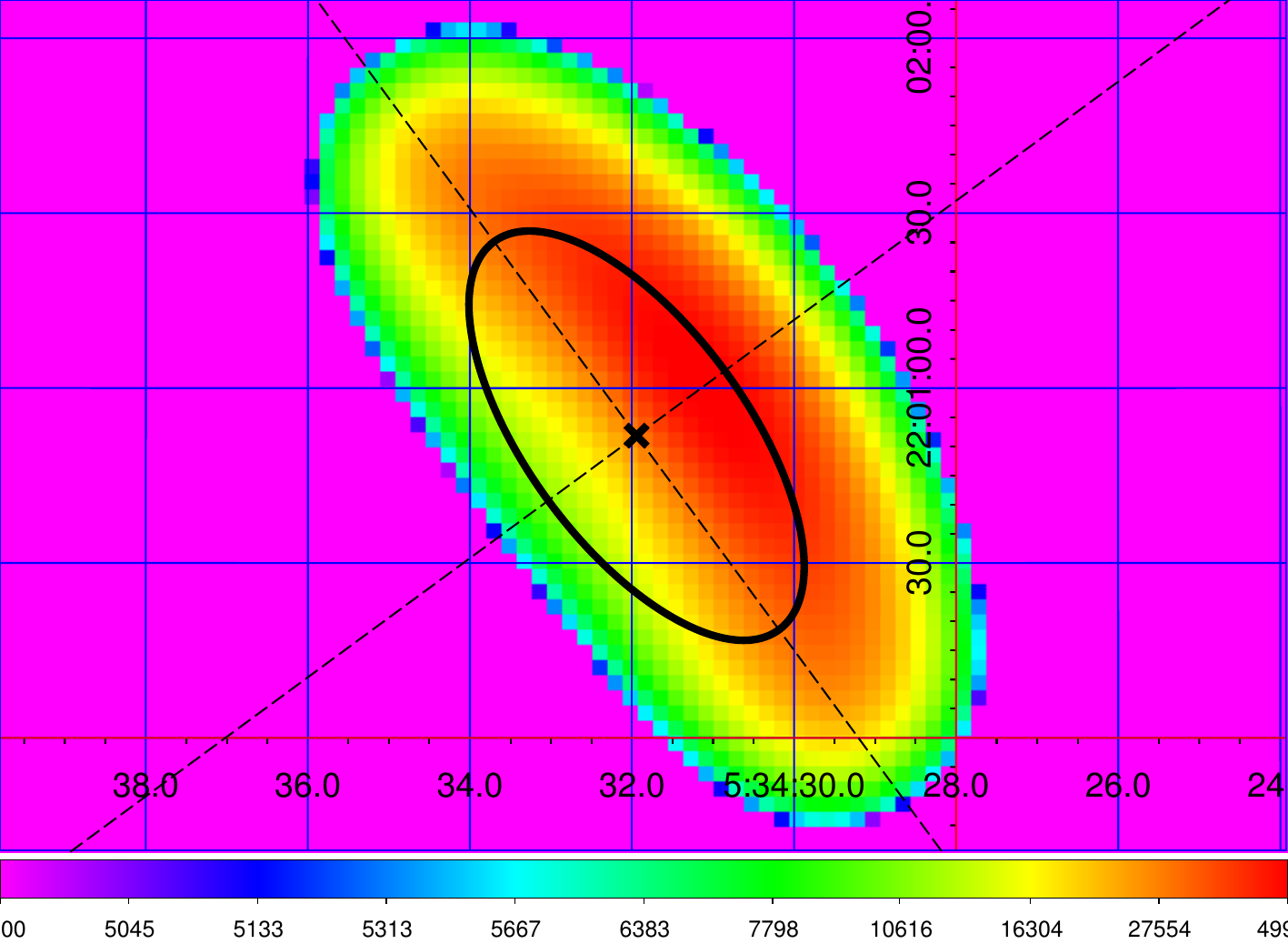}
   \subcaption{}
  \end{minipage}
  \begin{minipage}{0.5\textwidth}
   \centering
   \includegraphics[width=\textwidth]{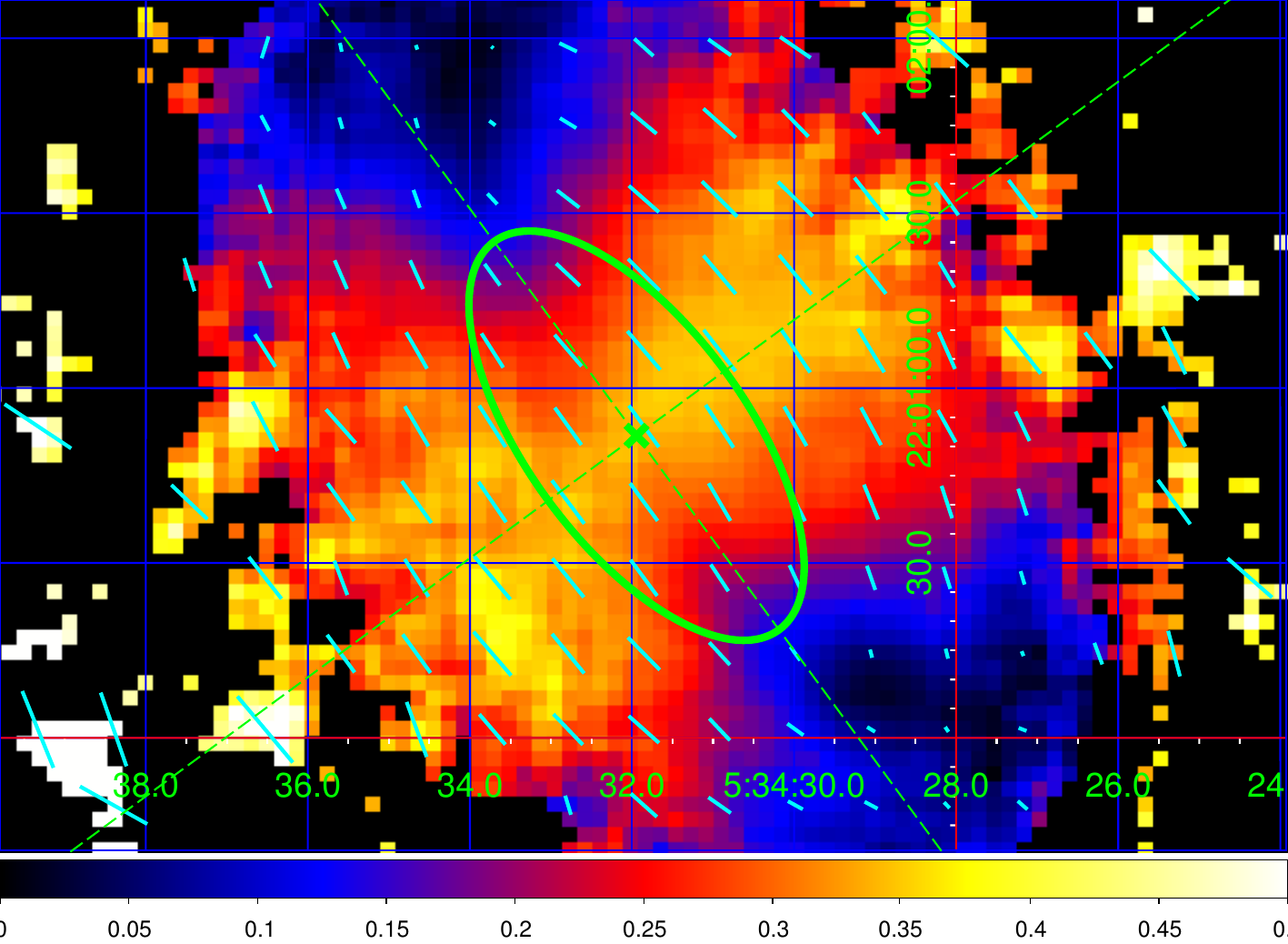}
   \subcaption{}
  \end{minipage}
 \end{tabular}
 \caption{Smoothed (a) Stokes $I$ map (in unit of counts) and (b) PD map of the simulated Crab PWN
 convolved with the IXPE response in 2--8~keV, both in J2000.0 equatorial coordinate. Like figure~1(c), only the pixels with a significance of more than 3$\sigma$ and ${\rm MDP_{99}<100\%}$ or those
with ${\rm MDP_{99}}<20\%$ are shown in the PD map.
The PSR position, spine of the X-ray torus and its axes \citep{Ng2004} are indicated for reference. 
The reconstructed magnetic field directions are overlaid  in cyan segments in the PD map.}\label{.....}
\end{figure}

\begin{figure}
 \begin{tabular}{cc}
  \begin{minipage}{0.5\textwidth}
   \centering
   \includegraphics[width=\textwidth]{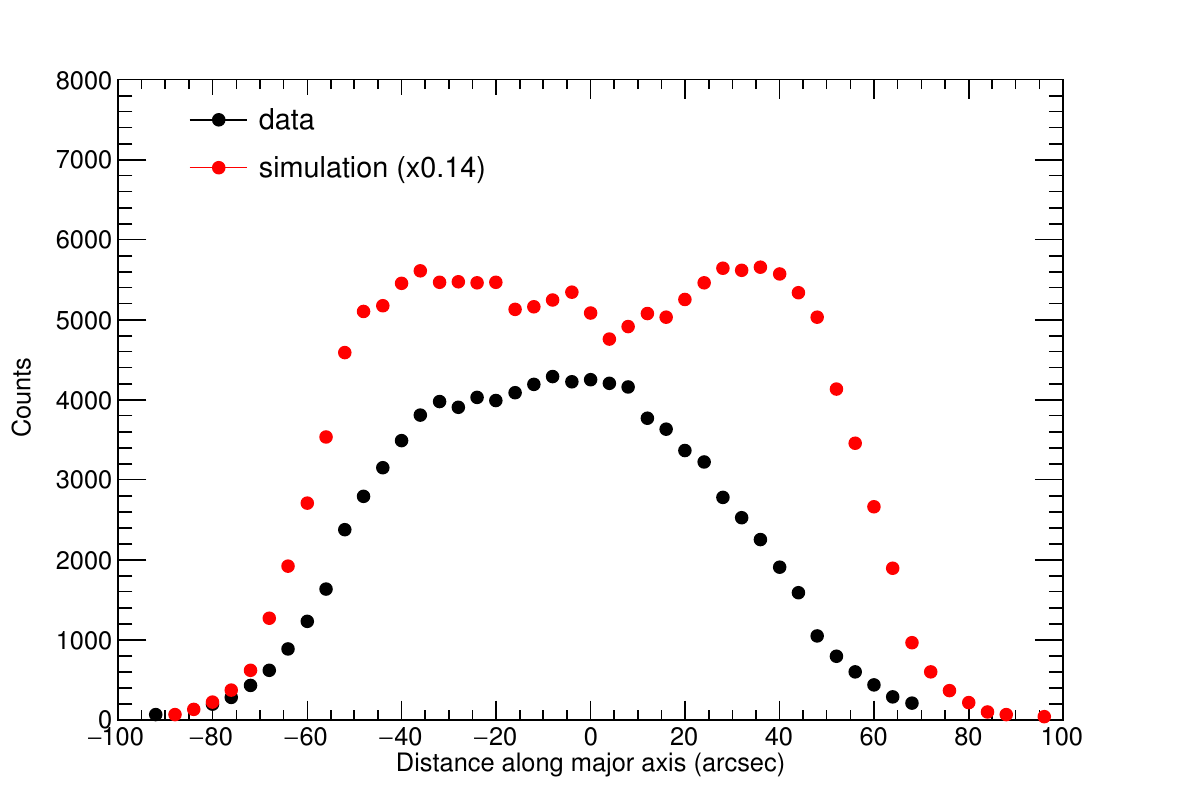}
   \subcaption{}
  \end{minipage}
  \begin{minipage}{0.5\textwidth}
   \centering
   \includegraphics[width=\textwidth]{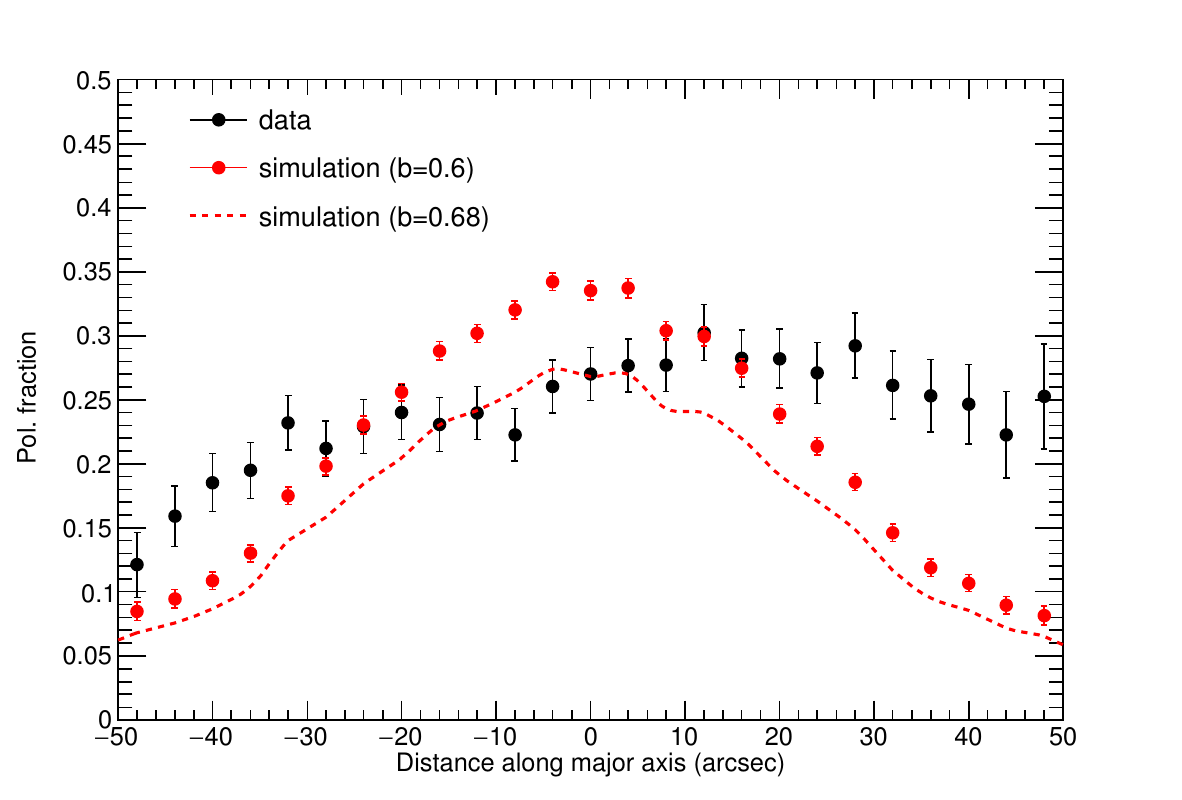}
   \subcaption{}
  \end{minipage} \\
  \begin{minipage}{0.5\textwidth}
   \centering
   \includegraphics[width=\textwidth]{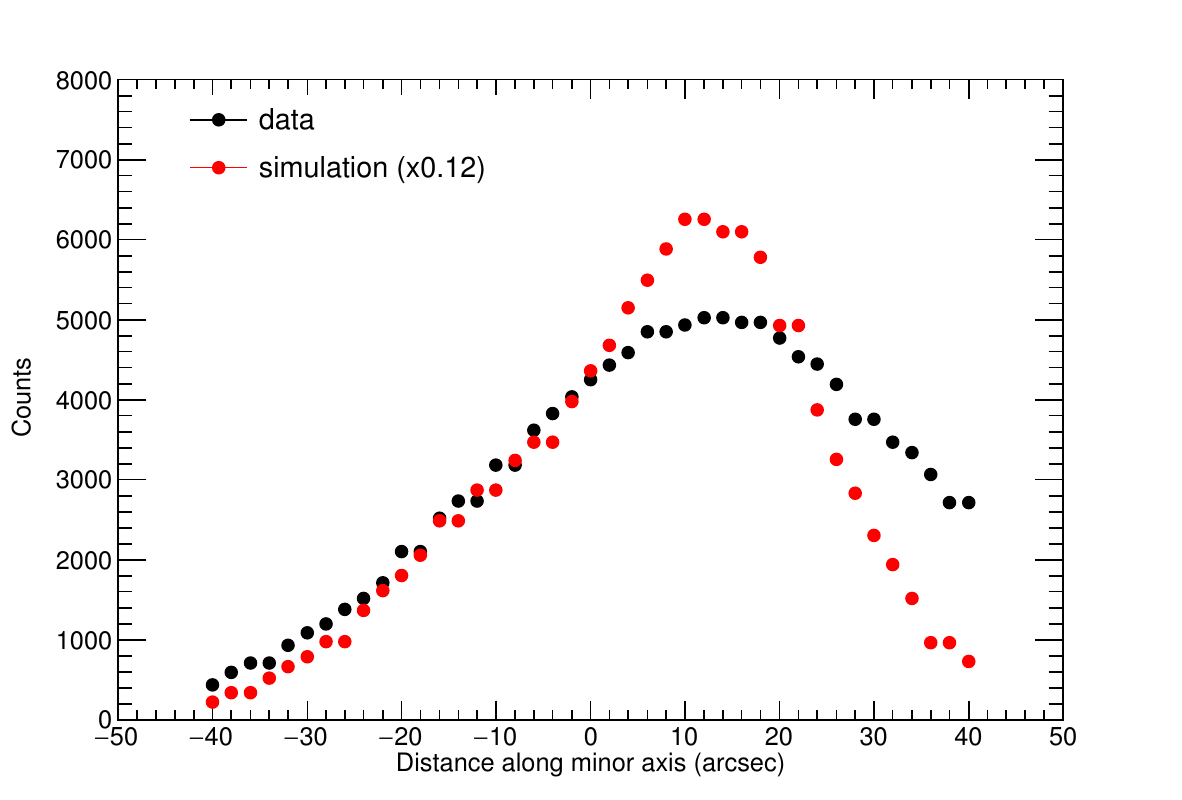}
   \subcaption{}
  \end{minipage}
  \begin{minipage}{0.5\textwidth}
   \centering
   \includegraphics[width=\textwidth]{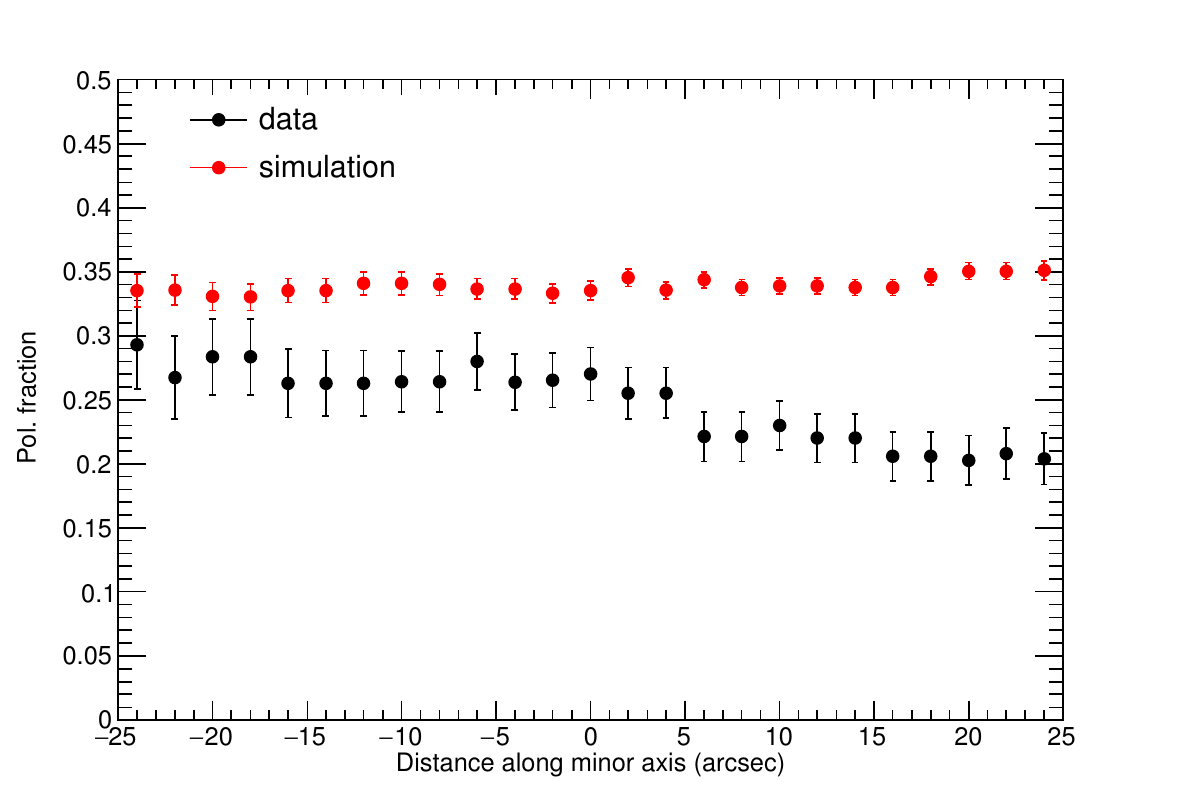}
   \subcaption{}
  \end{minipage}
 \end{tabular}
 \caption{Comparison of the observed and simulated profiles of the Crab PWN along the  (upper row) major and  (lower row) minor axes.
The profiles of the count and PD are shown in the left and right panels, respectively.
 The vertical scales for the simulation are adjusted arbitrarily in the count profiles.
See  text for detail of the simulation.
}\label{.....}
\end{figure}

\subsection{Polarization Properties of Jets}

As described in section~3.1, 
moderately low PD areas are seen at high latitudes along the axis of the jets
seen in the Chandra image.
The area corresponding to the southern jet is sandwiched between high PD areas. The same would be true for the norhtern jet, if the PD 
of an area around the dense filament were high (see also section~4.2).
We speculate that the jets have different PA and/or lower PD than regions outside the PSR spin axis
and produce these moderately low PD areas.
To investigate  the possibility,
we evaluate the  polarizations of the northern jet using source and off-source regions along a line
parallel to the major axis of the X-ray torus by \citet{Ng2004}, as shown in figure~5.
The line is $50^{"}$ away from the torus major axis,
and the source region (jet plus foreground and background emission along the line of sight) is  located on the projected torus axis.
The off-source region is defined in a high PD area
\timeform{26.3"}
away from the source toward the northeast along the line, where the offset value is comparable to the HPD.
The spatial distribution of the torus-like emission at the high latitude, which is the 
contamination to the jet emission in the source region, is unknown.
We examined the count rate profile of the Chandra data along the line $30^{"}$ away from the torus major axis,
where the contribution from the jet is small, 
and found that it is uniform within $\pm 15\%$ in the regions ${\le}30^{"}$ toward the northeast from the projected torus axis.
Hence, we  consider that the off-source region we defined is adequate to evaluate the polarization properties of the jet.
We also tested two other off-source regions, \timeform{21.9"} and \timeform{30.7"} away from the source region,
to evaluate the systematic uncertainty.
The size of each region  is $5 \times 5$ pixels. We calculate the background-subtracted Stokes $I$, $Q$, and $U$ and derive the PD and PA.
Table~1 tabulates the derived values of PD and PA.  
We also attempted a similar analysis to evaluate the  potential polarization  in the southern jet, 
but found the source-to-background count-ratio to be small
(${\le}1.15$), yielding a result with a meaninglessly large 1$\sigma$ error of the PD to be ${\ge}$40\%.

To summarize,
we found that the northern jet should have a PD ${\sim}$30\% and PA ${\sim}120^{\circ}$, with a detection significance of ${\sim}2{\sigma}$,
in order to justify the lower polarization along the projected torus axis.
 In comparison, the northern high-PD area gives PD${\sim}45\%$ and PA${\sim}160^{\circ}$.
Therefore, although the systematic and statistical errors are large, 
the estimated PA of the jet is roughly parallel to the observed direction of the jet,  
i.e., the magnetic field direction is perpendicular to it,
as expected in the situation where the magnetic field is compressed by the jet.
This possible superposition of polarized radiation from the jet,
with different properties than 
regions outside the PSR spin axis,
may produce moderately low PD areas through depolarization of the (originally) higher PD 
and different PA along the line of sight
at the high latitude.

\begin{table}
\tbl{Estimated polarization of the northern jet with three separate off-source (BG) regions}{
\begin{tabular}{cccc}
\hline
 & BG1 (closest) & BG2 & BG3 (farthest) \\
\hline \hline
PD & $(38.8\pm20.5)\%$ & $(26.9\pm10.8)\%$ & $(18.4\pm9.1)\%$ \\
PA (deg) & $117\pm15$ & $122\pm11$ & $134\pm14$ \\
S/BG ratio & 1.25 & 1.59 & 1.75 \\
\hline
\end{tabular}}
\end{table}

\section{Summary}
We carried out a detailed analysis of the spatially-resolved X-ray polarization data of the Crab PWN obtained with IXPE, following the  initial analysis report by \citet{IXPE-Crab}.
We  investigated how the polarization properties develop, depending on  the distance from the PSR.
We found that the reconstructed magnetic field was parallel to the pulsar wind in the inner region close to the PSR 
and then it gradually deviated to  the
east-west direction as a function of the distance from the PSR. 
We also found that a low-PD area to the west of the X-ray torus is not due to
the steep spatial variation of the PA.
Instead, the presence of a dense filament seen in the optical band indicates that
the flow of the pulsar wind is deflected and this produces variations in the magnetic field direction
leading to the low-PD region.
We employed a phenomenological model of the X-ray torus with simplistic toroidal and turbulent magnetic fields
 and obtained the randomness parameter $b$ to be ${\sim}$2/3
through  comparison between the data and model.
We also investigated possible jet polarizations. Although the errors are large,
the northern jet seems to have a PD of ${\sim}$30\%, which is 
 much lower than the high PD  observed  at the high latitude and PA of ${\sim}120^{\circ}$, 
which differs  from the PA  at the high latitude by ${\sim}40^{\circ}$.

\bigskip
\begin{ack}
The Imaging X-ray Polarimetry Explorer (IXPE) is a joint US and
Italian mission. The US contribution is supported by the National
Aeronautics and Space Administration (NASA) and led and managed
by its Marshall Space Flight Center (MSFC), with industry partner
Ball Aerospace (contract NNM15AA18C).

The Italian contribution is supported by the Italian Space Agency
(Agenzia Spaziale Italiana, ASI) through contract ASI-OHBI-2017-
12-I.0, agreements ASI-INAF-2017-12-H0 and ASI-INFN-2017.13-
H0, and its Space Science Data Center (SSDC) with agreements
ASI-INAF-2022-14-HH.0 and ASI-INFN 2021-43-HH.0, and by the
Istituto Nazionale di Astrofisica (INAF) and the Istituto Nazionale
di Fisica Nucleare (INFN) in Italy.
This research used data products provided by the IXPE Team
(MSFC, SSDC, INAF, and INFN) and distributed with additional
software tools by the High-Energy Astrophysics Science Archive
Research Center (HEASARC), at NASA Goddard Space Flight
Center (GSFC).

This research used data products provided by the IXPE Team
(MSFC, SSDC, INAF, and INFN) and distributed with additional
software tools by the High-Energy Astrophysics Science Archive
Research Center (HEASARC), at NASA Goddard Space Flight
Center (GSFC).
This work was supported by JSPS KAKENHI Grant Number 	
23H01186 (T.M.), 22K14068 (E.W.), 19H00696 (S.G.), and 	22K03681 (S.S.).
N.B. was supported by the INAF MiniGrant "PWNnumpol - Numerical Studies of Pulsar Wind Nebulae in The Light of IXPE".

\end{ack}

\appendix
\section{Polarization Analysis  Formulation}
The event-by-event Stokes parameters are given  by
$i_{k} \equiv 1$, $q_{k} \equiv 2 \cos2\phi_{k}$, and $u_{k} \equiv 2 \sin2\phi_{k}$,  where $\phi_{k}$
 is the reconstructed photoelectron direction of the event number $k$. 
 Stokes $I$, $Q$, and $U$ of the source  are then estimated  with
$I=\Sigma i_{k} = \Sigma 1$, $Q=\Sigma q_{k}/\mu$, and $U=\Sigma u_{k}/\mu$, respectively,
 providing that the effective area and modulation factor ($\mu$) are energy independent.
The variance $V(q)$ of  Stokes $Q$ of  an event is given  by $V(q) = \langle q^{2} \rangle - \langle q \rangle^{2}$.
 Assuming that the source polarization and/or $\mu$  are significantly smaller than 1, which is usually the case, the relation 
$\langle q^{2} \rangle \gg \langle q \rangle^{2}$ holds and hence $V(q) \approx \langle q^{2} \rangle =\frac{1}{N}\Sigma (2 \cos2\phi_{k})^{2}$,
where $N$ is the number of events.
 Then, we can estimate $V(Q)$  with $V(Q)=NV(q)/\mu^{2} \approx \Sigma (2\cos2\phi_{k})^{2}/\mu^{2}$.
 Similarly, we obtain $V(U) \approx \Sigma (2\sin2\phi_{k})^{2}/\mu^{2}$.
The source polarization  parameters are calculated as 
${\rm PD}=\sqrt{Q^{2}+U^{2}}/I$ and ${\rm PA}=\frac{1}{2}\arctan(U/Q)$.
When the number of events is large as is often the case,
$\sqrt{V(I)}/I=1/\sqrt{N}$ is much smaller than $\sqrt{V(Q)}/Q$ and $\sqrt{V(U)}/U$, and  the errors of PD and PA
 practically depend only on $V(Q)$ and $V(U)$.
To calculate ${\rm MDP_{99}}$, we  consider the statistical error of  a zero polarization source;
 $V(Q) \approx \Sigma(2\cos2\phi_{k})^{2}/\mu^{2} = 4(N/2)/\mu^{2}=2N/\mu^{2}$ (because $\langle \cos^{2}\phi \rangle = 1/2$)
and $V(U) \approx 2N/\mu^{2}$.
This  yields $\sigma(Q/I)=\sqrt{V(Q)}/N=\sqrt{2/N}/\mu$ and $\sigma(U/I)=\sqrt{2/N}/\mu$, resulting in
${\rm MDP_{99}}=3\sqrt{2/N}/\mu=4.29/\mu\sqrt{N}$.

In a realistic case, the modulation factor $\mu$ is energy dependent. Denoting it as $\mu_{k}$, we  define
$I_{\rm corr} \equiv \Sigma i_{k}=N$ (not changed), $Q_{\rm corr} \equiv \Sigma q_{k}/\mu_{k}$, and $U_{\rm corr} \equiv \Sigma u_{k}/\mu_{k}$.
The effective area is also energy dependent in most cases ($A_{k}$).
Stokes parameters  are given by 
\begin{eqnarray}
\tilde{I}_{\rm corr} & \equiv & \Sigma i_{k}/A_{k} (\equiv \tilde{N}), \\
\tilde{Q}_{\rm corr} & \equiv & \Sigma q_{k}/\mu_{k}/A_{k}, \\
\tilde{U}_{\rm corr} & \equiv & \Sigma u_{k}/\mu_{k}/A_{k}.
\end{eqnarray}
The source polarization  parameters are then obtained with 
\begin{eqnarray}
{\rm PD} & = & \sqrt{\tilde{Q}_{\rm corr}^{2}+\tilde{U}_{\rm corr}^{2}}/\tilde{I}, \\
{\rm PA} & = & \frac{1}{2}\arctan(\tilde{U}_{\rm corr}/\tilde{Q}_{\rm corr}).
\end{eqnarray}
By taking account of the error propagation, we obtain
\begin{eqnarray}
V(\tilde{Q}_{\rm corr}) & = & \{ (1/\mu_{1})^{2}(1/A_{1})^{2} + \cdots + (1/\mu_{N})^{2}(1/A_{N})^{2} \} V(q) \nonumber \\
& = & \Sigma (1/\mu_{k})^{2}(1/A_{k})^2 \cdot \Sigma q_{k}^{2}/N, \\
V(\tilde{U}_{\rm corr}) & = & \Sigma (1/\mu_{k})^{2}(1/A_{k})^2 \cdot \Sigma u_{k}^{2}/N.
\end{eqnarray}
Since $\sqrt{V(\tilde{I}_{\rm corr})}/\tilde{I}_{\rm corr}$ is much smaller than
$\sqrt{V(\tilde{Q}_{\rm corr})}/\tilde{Q}_{\rm corr}$ (and $\sqrt{V(\tilde{U}_{\rm corr})}/\tilde{U}_{\rm corr}$),
we can calculate the errors of the PD and PA by using
$V(\tilde{Q}_{\rm corr})$ and $V(\tilde{U}_{\rm corr})$ only.
In particular, when the polarization significance is high,
both the PD and PA distributions are symmetric, and their 1${\sigma}$ errors  are
\begin{eqnarray}
\frac{\sigma({\rm PD})}{\rm PD} & \approx & 
\frac{\sqrt{\tilde{Q}_{\rm corr}^{2}V(\tilde{Q}_{\rm corr})+
\tilde{U}_{\rm corr}^{2}V(\tilde{U}_{\rm corr})}}
{\tilde{Q}_{\rm corr}^{2}+\tilde{U}_{\rm corr}^{2}}, \\
\sigma({\rm PA}) & \approx & \frac{1}{2}\frac{\sigma({\rm PD})}{\rm PD}.
\end{eqnarray}
To calculate ${\rm MDP_{99}}$, we consider the statistical error of  a zero polarization source and obtain
$V(\tilde{Q}_{\rm corr}) \approx \Sigma (1/\mu_{k})^{2}(1/A_{k})^2 \cdot \Sigma q_{k}^{2}/N = \Sigma(1/\mu_{k})^{2}(1/A_{k})^{2} \cdot 4 (N/2)/N = 
2\Sigma(1/\mu_{k})^{2}(1/A_{k})^{2}$.  Similarly, $V(\tilde{Q}_{\rm corr}) \approx 2\Sigma(1/\mu_{k})^{2}(1/A_{k})^{2}$.
 Defining  $N \langle \frac{1}{\mu^{2}}\frac{1}{A^{2}} \rangle \equiv (\frac{1}{\mu_{1}} \frac{1}{A_{1}})^{2} + \cdots + (\frac{1}{\mu_{N}} \frac{1}{A_{N}})^{2}$, 
we have
$\sigma(\frac{\tilde{Q}_{\rm corr}}{\tilde{I}_{\rm corr}}) = \sigma(\frac{\tilde{U}_{\rm corr}}{\tilde{I}_{\rm corr}}) = 
\frac{\sqrt{2N}}{\tilde{N}}\sqrt{\langle \frac{1}{\mu^{2}}\frac{1}{A^{2}} \rangle}$.
 Finally, we obtain
\begin{equation}
{\rm MDP_{99}}=\frac{3\sqrt{2N}}{\tilde{N}}\sqrt{\langle \frac{1}{\mu^{2}}\frac{1}{A^{2}} \rangle} = 
\frac{4.29}{\tilde{N}/\sqrt{N}}\sqrt{\langle \frac{1}{\mu^{2}}\frac{1}{A^{2}} \rangle}.
\end{equation}

\section{Crab PWN Maps with Normal Binning}
While the sliding-box smoothing maintains fine resolution and allows us to study polarization properties in detail,
the statistical errors of adjacent pixels are not independent. For completeness, we also show the
Stokes $I$ and PD maps of the Crab PWN in figure~8, in which normal binning of 
$5 \times 5$ pixels is applied.

\begin{figure}
 \begin{tabular}{cc}
  \begin{minipage}{0.5\textwidth}
   \centering
   \includegraphics[width=\textwidth]{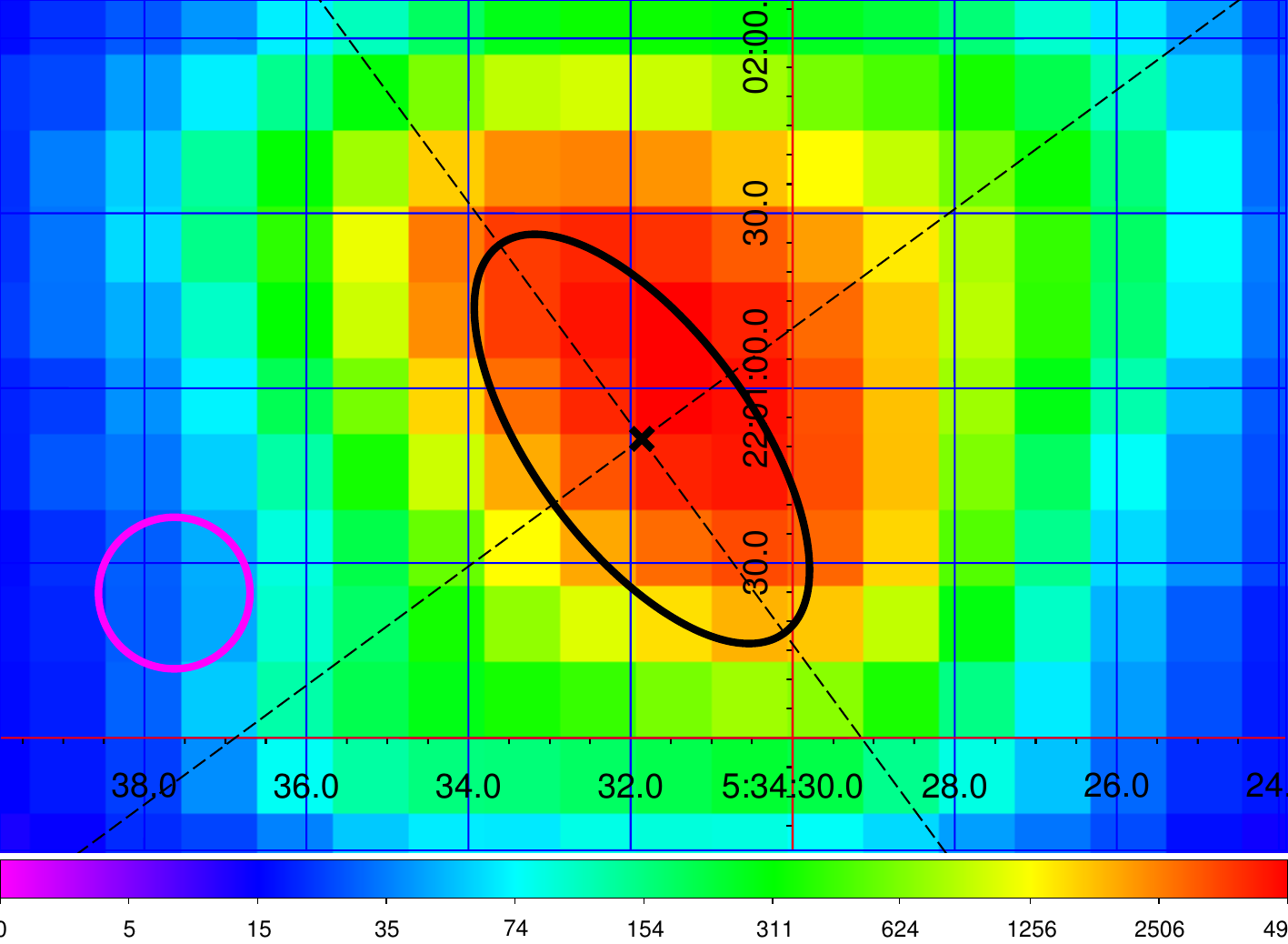}
   \subcaption{}
  \end{minipage}
  \begin{minipage}{0.5\textwidth}
   \centering
   \includegraphics[width=\textwidth]{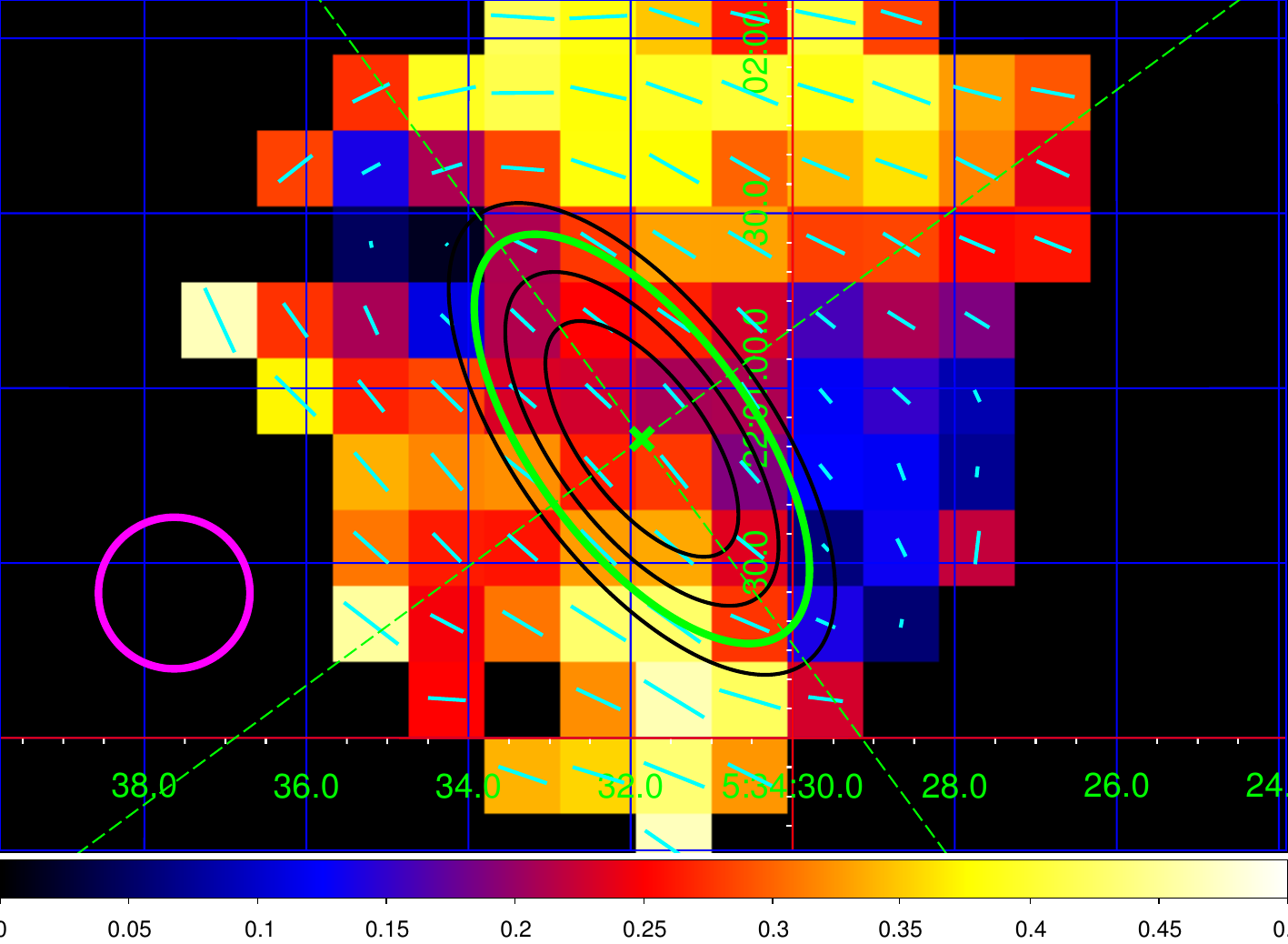}
   \subcaption{}
  \end{minipage}
 \end{tabular}
 \caption{
(a) Stokes $I$ map (in unit of counts) and (b) PD map of the Crab PWN obtained with IXPE in 2--8~keV,
both in J2000.0 equatorial coordinate. They are basically the same as figures 1(b) and 1(c), respectively,
but normal binning of $5 \times 5$ pixels is applied.
}\label{.....}
\end{figure}

\end{document}